\documentclass[]{jfm}

\usepackage{graphicx}
\usepackage{siunitx}
\usepackage{newtxtext}
\usepackage{newtxmath}
\usepackage{natbib}
\usepackage{tikz,pgfplots}
\usetikzlibrary{shapes.geometric}

\usepackage[hypertexnames=false]{hyperref}
\usepackage{xcolor}
    \definecolor{myRed}{RGB}{255, 40, 60}
	\definecolor{myActualBlue}{RGB}{25, 165, 180}
    \definecolor{myOrange}{RGB}{255, 153, 102}

\hypersetup{
    colorlinks = true,
    urlcolor   = blue,
    citecolor  = black,
}

\newcommand{\RomanNumeralCaps}[1]
\linenumbers

\newcommand{\tinySpace}{\hspace{0.25pt}}
\newcommand{\miniSpace}{\hspace{0.5pt}}
\newcommand{\microSpace}{\hspace{1pt}}

\renewcommand{\eg}{\textit{e.g.}}
\newcommand{\ie}{\textit{i.e.}}

\newcommand{\Pe}{\text{Pe}}

\newcommand{\rdrop}{r_{\rm d}}

\newcommand{\tdrip}{t_{\rm{0}}}

        \newcommand{\dr}{\hfilm_{r}}

        \renewcommand{\sf}{\psi}
        
        \newcommand{\vold}{V_{\rm d}}

    \newcommand{\tMax}{t_{\rm M}}

    \usepackage{cancel}

\newcommand{\smelv}{\eta}

    \newcommand{\Rsm}{R}

    \newcommand{\shapeFac}{\Psi}

    \newcommand{\xiVec}{\textbf{e}_{\xi}}
    \newcommand{\zetaVec}{\textbf{e}_{\zeta}}
        
    \newcommand{\rnought}{r_{0}}
    \newcommand{\hnought}{\hfilm_{0}}
    \newcommand{\hun}{\hfilm_{1}}
    
    \newcommand{\dt}{\partial_{t}}

    \renewcommand{\dr}{\partial_{r}}

        \newcommand{\dzetint}{\,\mathrm{d}\zeta}

    \newcommand{\dxi}{\partial_{\xi}}
    \newcommand{\dxxi}{\partial_{\xi\xi}}

    \newcommand{\dze}{\partial_{\zeta}}
    \newcommand{\dzze}{\partial_{\zeta\zeta}}

    \newcommand{\unDemi}{\raisebox{0.625pt}{\scriptsize{1}}\hspace{-1pt}\raisebox{0.75pt}{\scriptsize{/}}\hspace{-0.5pt}\raisebox{-0.75pt}{\scriptsize{2}}} 

    \newcommand{\pinf}{p_{\infty}}
    \renewcommand{\u}{\mathbf{u}}
    \newcommand{\q}{\mathbf{q}}

    \newcommand{\flux}{q}

    \newcommand{\hfilm}{h}

        \newcommand{\rStnry}{\ell_{\text{sr}}} 
        
            \newcommand{\lFront}{\ell_{\text{f}}}
            
        \newcommand{\rFront}{\ell_{\text{f}}} 
        \newcommand{\xiFront}{\ell_{\text{f}}} 

    \newcommand{\hfilmStnry}{h_{\text{s} }}
    \newcommand{\hfilmStnryZ}{h_{\text{s,0} }}

    \newcommand{\hdroplets}{\mathcal{H}_{\rm d}}
    \newcommand{\hdrop}{h_{\rm d}}
    \newcommand{\qdrain}{q_{\rm d}}

    \newcommand{\smAng}{\varphi}
    \newcommand{\phiStar}{\varphi_{\raisebox{-1.65pt}{\hspace{-1.65pt}\scriptsize{$\star$}}}}

    \newcommand{\Vdrop}{V_{\rm d}}

    \newcommand{\Dt}{\Delta t}
    
    \newcommand{\Dxi}{\Delta \xi}

    \newcommand{\ndrip}{n_0}
    
\newcommand{\tdrain}{t_{\mathrm{d}}}

    \newcommand{\qin}{\textit{q}_{\hspace{1pt}\text{in}}}

    \newcommand{\Qdrip}{\textit{Q}_{\hspace{1pt}\text{drip}}}
    \newcommand{\Qdrain}{\textit{Q}_{\hspace{1pt}\text{drain}}}

\title[Thin film drainage on a stalagmite]{Discrete inflow and drainage dynamics of a thin film over a stalagmite of variable shape}

\author[J. Parmentier, V. E. Terrapon and T. Gilet]{Justine Parmentier\aff{1}
  \corresp{\email{justine.parmentier@gmail.com}},
  Vincent E. Terrapon\aff{2}
 and Tristan Gilet\aff{2}}

\affiliation{\aff{1}The Njord Centre, University of Oslo, Oslo, Norway
\aff{2}Department of Aerospace and Mechanical Engineering, University of Liège, Liège, Belgium}

\begin{document}
\maketitle

\begin{abstract}
Stalagmites in karstic caves preserve valuable palaeoclimate records through calcium-rich layered deposits, presenting curvature variations both across and within individual stalagmites. Stalagmites always remain covered by a thin water film fed by a discrete inflow of drops, which bring in new ions in solution for the stalagmites to grow. However, the gravity-induced drainage of this film and its response to the stalagmite underneath shape and the discrete drop inflow remain poorly characterised in existing growth models. To address these limitations, we develop a theoretical framework that captures the combined effects of shape curvature and discrete drop inflow on thin film drainage dynamics, starting from Reynolds lubrication theory expressed in curvilinear coordinates. From there, we show that the limiting cases of thickness-dominated and inclination-dominated drainage translate into distinct scaling laws for both the front propagation position and stationary film thickness. We further validate these results by numerically solving the governing equations. Finally, experimental measurements conducted in both cave and lab settings confirm the predicted stationary film thickness. Our findings provide insights into the influence of substrate shape and inflow dynamics on thin film drainage, with implications for stalagmite growth modelling and other gravity-driven surface flows.
\end{abstract}


\section{Introduction}

Palaeoclimatology refers to the study of past natural climate variations, as opposed to human-induced climate change~(\eg, \cite{Giec2022}). A wide range of palaeoclimate proxies record past temperature and precipitation variations (\eg, \cite{PetitPaleoBook}), refining our understanding of long-term climate patterns and informing future predictions. Among them, stalagmites in karstic caves are particularly valuable, especially in regions lacking glacial ice cores or other continental proxies (\eg, \cite{Wang2008}). Their significance stems from (i) their global distribution (\eg, \cite{Goldscheider2020}), (ii) their long-term preservation given the stability of environmental conditions (temperature, atmosphere composition) inside caves (\eg, \cite{Burns2002}) and (iii) the presence of a variety of palaeoclimate indicators, such as trace elements and isotopic ratios, that can be precisely dated (\eg, \cite{Taylor1987, Tan2006}). 
Specifically, the laminae revealed in a stalagmite cross-sectional cut are correlated to the past upstream flows and soil coverage above the cave (\eg, \cite{XU2015559, Rossi2016}). Hence, the shape of these laminae, and therefore of the stalagmite top surface, changes over time in response to variable conditions (\eg, \cite{DreybrodtBook, Baldini2021, Baker2022}). As a result, variations in stalagmite composition can be directly linked to external climate changes, such as shifts in precipitation or temperature, rather than local environmental noise. 

Yet, the causal link between past flows and stalagmite shape variability remains poorly understood. Existing models describing stalagmite growth involve strong simplifying assumptions regarding the aerodynamics and hydrodynamics of drops impacting stalagmites in caves (\eg, \cite{DreybrodtBook, Hansen2007}). For instance, drops were assumed to always land at the apex of the stalagmite, thereby feeding the thin residual film covering it in one central point. However, it was recently shown that the falling drop interacts with its own wake. Hence, its impact position is sometimes scattered over several centimetres, which largely determines stalagmite width (\eg, \cite{Parmentier2019}). The concave shape exhibited by some stalagmites, called drip cups (\eg, \cite{BreitenbachEGU2024}), was also associated with drops splashing at impact. However, most drop impacts in caves lead to splashing and cannot, therefore, be related to a particular stalagmite shape (\eg, \cite{Cossali2004}). Another assumption of existing stalagmite growth models is that the thin residual film lying on top of the stalagmite remains uniform in time and space. This overlooks the possible interplay between the intermittent inflow of discrete drops feeding the film and the effect of stalagmite shape on the film thickness. Hence, this assumption may not always be valid to describe stalagmite morphogenesis. 

Due to the hydrophilicity of calcite, stalagmites are naturally covered by very thin films ($\sim \SI{100}{\micro\meter}$) in comparison to their average width (between \SI{5}{\centi\meter} and \SI{50}{\centi\meter}) or the more or less constant radius of the falling drops (approximately \SI{2.7}{\milli\meter}, which is the capillary length of water in air, \eg, \cite{Parmentier2019}). Some exceptions include drip cups (\eg, \cite{BreitenbachEGU2024}), which can form pools of water up to \SI{1}{\centi\meter} deep due to their unique concave shapes.  Thin film flows are ubiquitous in both natural and industrial settings. They can be categorised into (i) free surface thin film flows such as lava flows (\eg, \cite{Dietterich2022,Griffiths2000}) or free soap film drainage (\eg, \cite{Barigou1994}), and (ii) flows between confined boundaries like synovial flows in joints (\eg, \cite{Maqbool2022}), Bretherton bubbles in narrow tubes (\eg, \cite{Bretherton1961}) or oil and gas transport in pipes (\eg, \cite{Zheng2015}). We focus exclusively on free-surface thin film flows here, whose dynamics responds primarily to competing viscous and gravity effects. These flows can be induced by the spreading of a given liquid volume onto a substrate, or by a constant flow feeding it (\eg, \cite{Huppert1982, Simpson1997}). In both cases, the resulting thin film dynamics depends strongly on the boundary conditions considered, substrate geometry or other competing physical processes at play such as the deformability of the substrate (\eg, \cite{Spence1985}, \cite{Lister2012}), its permeability which induces partial fluid loss (\eg, \cite{Zheng2015}, \cite{Zheng2022}), or even inertial effects (\eg, \cite{Momen2017}). 

In many configurations, self-similar solutions can be derived (\eg, \cite{Huppert1982,Gratton1990,Jensen1994}), where the film thickness and its front position can both be expressed as power laws of time (\eg, \cite{Zheng2022}). However, most of these models assume idealised or particularised geometries (flat or linear slopes) and continuous inflow conditions. In contrast, natural cave environments involve non-trivial substrate geometries and discrete drop impacts. Intermittent discrete inflows feeding the film are usually not taken into account in modelling thin film drainage (\eg, \cite{Simpson1997, Zheng2022}), although they could be of interest for geophysical or industrial chemical applications, such as trickle-bed reactors (\eg,~\cite{Guo2022}). To our knowledge, there is no dedicated general model describing the film as a function of the shape of the substrate underneath, nor the effect of a discrete inflow on subsequent film drainage dynamics. This raises the following question: how does a thin film fed by discrete drops evolve over a curved stalagmite?

In this study, we thus model how the underneath stalagmite shape and the discrete drop inflow jointly affect the spatio-temporal response of the film to the competing gravity and viscosity. Using Reynolds lubrication theory expressed in a curvilinear framework, we derive in Sec.~\ref{sec:theory} the governing equations of gravity-driven drainage on an axisymmetric stalagmite of general shape. In Sec.~\ref{subsec:num-examples}, we solve the equations numerically in idealised flat and parabolic geometries, which allows us to explore the transient and stationary dynamics of the film under discrete drop inflow. We then extend our theoretical analysis to limiting cases of flat and parabolic stalagmite profiles in Sec.~\ref{subsec:scalings}. We further compare these predictions to experimental film thickness measurements taken both in caves and in a controlled lab setting in Sec.~\ref{sec:exp}, using complementary experimental techniques. We conclude by discussing in Sec.~\ref{sec:discussion} the interplay between inflow, geometry, and film thickness, and outline implications for natural cave formations. This work combines theoretical modelling, numerical resolution of the equations and experimental measurements that, altogether, highlight how the local and global film thickness dynamics may affect subsequent stalagmite growth.

\begin{figure}
\centerline{\includegraphics[width = \linewidth]{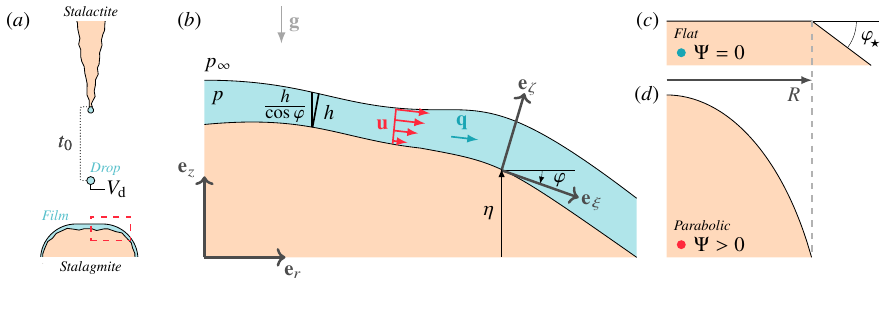}}
    \caption{(a) General phenomenology: a drop of volume~$\Vdrop$ is shown just before detaching from a stalactite hanging from the cave ceiling. The previous drop fell~$\tdrip$ earlier (the dripping period) and contributed to the thin film covering the underlying stalagmite. Successive drops from the same stalactite form a thin film on the stalagmite surface. The dashed red rectangle indicates the region magnified in (b). (b) Cross-sectional sketch at a fixed polar angle of an axisymmetric stalagmite (in beige), covered by a thin film of local thickness~$\hfilm(\xi)$ (in blue). At position~$(r, z)$, the stalagmite elevation is~$\smelv(r)$, and the local inclination angle with respect to the horizontal is~$\smAng(\xi)$ (positive when downward). A local curvilinear coordinate system~$(\xi, \zeta)$ is defined accordingly. The pressure field in the film is~$p(\xi, \zeta)$, with atmospheric pressure~$\pinf$, and the local velocity field is~$\u(\xi, \zeta)$, with integrated cross-sectional flux~$\q(\xi)$. Gravity acts along~$\mathbf{g}$. (c) Example of a flat stalagmite ($\shapeFac = 0$), modelled as a disk of radius~$\Rsm$, surrounded by a cone of constant opening angle~$\phiStar$. (d) Example of a stalagmite with a parabolic profile of revolution ($\shapeFac > 0$), defined over the domain~$r \in [0, \Rsm]$ by Eq.~\eqref{eq:sm-shape}.}
    \label{fig:chap3-model-bigSchematic}
\end{figure}

\section{Theoretical description}
\label{sec:theory}

A stalagmite in a karstic cave grows upward by approximately $\SIrange{1}{1000}{\micro\meter}$ over a year 
(\eg, \cite{Baker2008}). It is formed by the progressive deposition of calcium carbonate brought by the successive water drops falling from the stalactite located directly above the corresponding stalagmite. In this study, we only focus on individual stalagmites associated with a single stalactite (see Fig.~\ref{fig:chap3-model-bigSchematic}\;(a)). Upon impact, each drop merges with the existing liquid layer, forming a thin film that spreads over the stalagmite surface. Due to the high humidity levels inside caves, this film persists and drains slowly under gravity, enabling mineral precipitation over time.

In this section, we describe the gravity-driven drainage of a thin liquid film lying on top of a stalagmite modelled as an axisymmetric body, a reasonable assumption for most stalagmites fed by a single drip (\eg, \cite{Szymczak2025,Parmentier2019}). As represented in Fig.~\ref{fig:chap3-model-bigSchematic}\;(b), the radial system of coordinates along a given polar angle,~$(r, z)$, is aligned with the stalagmite axis of symmetry. The stalagmite top is either flat (horizontal) or curved, and may present macroscopic irregularities with a radius of curvature down to \SI{1}{\milli\meter}. By contrast, the film thickness lying on top of the stalagmite is of the order of~$\SIrange{30}{300}{\micro\meter}$~(\eg, \cite{Parmentier2023}), which is much smaller than the aforementioned minimum radius of curvature of the stalagmite surface. We thus consider a curvilinear coordinate system~$\left( \xi, \zeta \right)$ aligned with the local stalagmite geometry at each point. As shown in Fig.~\ref{fig:chap3-model-bigSchematic}\;(b), $\xi$, resp.~$\zeta$, represents the direction parallel, resp. normal, to the stalagmite surface. The corresponding unit vectors~$\xiVec$ and~$\zetaVec$ are locally defined at each point of the surface, with~$\xiVec$ inclined by an angle~$\smAng$ with respect to the horizontal, counted as positive downward. The film thickness measured normally to the solid/liquid interface is~$\hfilm$. We further approximate the stalagmite elevation profile,~$\smelv$, as either entirely flat or represented by a parabola oriented downward. These two shapes are commonly found among stalagmites (\eg, \cite{Szymczak2025}) and are defined by
\begin{equation}
    \smelv(r) = - \shapeFac r^2 \, ,  \qquad 0 \le r \le \Rsm \, , \label{eq:sm-shape}
\end{equation}
where we denote $\shapeFac$ the shape factor with units of inverse length, and where~$\Rsm$ is either the domain width or, in the degenerate case $\shapeFac = 0$, the stalagmite top surface radius. We restrict our analysis to convex surfaces ($\shapeFac > 0$), although the developed framework remains valid even for mildly concave surfaces ($\shapeFac \lesssim 0$).

We limit our study to an incompressible Newtonian fluid of density~$\rho$ and kinematic viscosity~$\nu$. The ions in solution are assumed sufficiently diluted to neglect hydrodynamic interactions. Under the thin film approximation, velocity gradients within the film in the direction normal to the surface are much larger than those parallel to the surface. Additionally, inertial effects can be neglected if changes in velocity occur at a timescale $\gg \hfilm^{2}/\nu \sim \SI{e-2}{\second}$, ensuring that the flow remains laminar (\ie, the Reynolds number $\ll 1$). In this case, the flow is predominantly oriented along~$\xiVec$, with a velocity $\u = u(\zeta) \, \xiVec$. The thin film approximation also implies that second-order derivatives parallel to the stalagmite surface become of negligible order. Therefore, under these approximations and in the absence of drop inflow, the Navier-Stokes momentum equation reduces to a balance between pressure gradients, and viscous and gravitational forces. We can thus adapt Reynolds lubrication theory (\cite{Reynolds1886}) to quantify the drainage dynamics in our curvilinear coordinate system. 
Under the lubrication approximation, the projection onto~$\xiVec$ and~$\zetaVec$ of the momentum balance reads
\begin{subeqnarray}
    \dxi p &=& \mu \dzze u + \rho g \sin{\smAng} \, \text{,} \\
    \dze p &=& - \rho g \cos{\smAng} \, ,
    \label{eq:intermediate}
\end{subeqnarray}
where $\mathbf{g} = -g \mathbf{e}_{z}$. 

To complete the well-posedness of the formulation, at the solid/liquid interface between the stalagmite and the film, we consider that a no-slip condition applies: $u(\zeta = 0) = 0$. The absence of significant normal stress at the liquid/air interface further yields $\dze u = 0$ and $p = \pinf$ in $\zeta = h$, with $\pinf$ the atmospheric pressure in the vicinity of the~stalagmite. As aforementioned, the stalagmite principal radii of curvature should be large enough to neglect the variations of the stalagmite elevation with the curvilinear abscissa. This translates here into~$\hfilm \ll 1/\left|\dxi \smAng\right|$. 
In addition, for an axisymmetric surface, the thin-film approximation requires the vicinity of the origin to be nearly flat, \ie, $\dxi \hfilm \rightarrow 0$ around $\xi = 0$. This ensures that both principal curvatures remain small compared to the inverse of the film thickness and that the thin film approximation remains valid. Pronounced curvature variations such as drip cups (\eg, \cite{BreitenbachEGU2024}), \ie, for which $\shapeFac \ll 0$, could yield the formation of liquid pools that violate the lubrication approximation. They are thus left out of this study.

If the stalagmite surface is also assumed  smooth enough, the Laplace pressure remains much smaller than the $\sim \SI{1}{\pascal}$ hydrostatic pressure within the film, so it can be neglected. The disjoining pressure can also be neglected; this pressure would arise between water and calcite from Van der Waals interactions and would only become comparable to the hydrostatic pressure for $\sim \SI{500}{\nano\meter}$-thick films (\eg, \cite{hamaker}). Inserting the boundary conditions for $\zeta = 0$ and $\zeta = h$ in Eqs.~\eqref{eq:intermediate} 
we eventually find that 
\begin{eqnarray} 
        p &=& \pinf + \rho g \left( \hfilm - \zeta \right) \cos{\smAng} \, \text{,} \label{eq:model-pressure} \\
        u &=& \dfrac{- g}{2 \nu} \zeta \left( 2h - \zeta \right) \left[ \left( \dxi \hfilm \right) \cos{\smAng} - \sin{\smAng} \right] \, . \label{eq:model-u-orig}  
\end{eqnarray}
The pressure at height~$\zeta$ in the film is thus primarily hydrostatic, but modulated by the inclination of the stalagmite surface underneath. Additionally, the half-parabolic profile of the velocity field is modulated by (i) the film thickness heterogeneity over the stalagmite, and (ii) the surface inclination.

The macroscopic evolution of the film can be connected to the above local fields by introducing the surface-parallel flux~$\q = q \xiVec$, which is obtained by integrating the velocity profile from Eq.~\eqref{eq:model-u-orig} across the film thickness normal to the solid/liquid interface. This flux represents the local drainage rate per unit width along the stalagmite surface and is expressed as
\begin{eqnarray}
    \flux = \int_{0}^{h}{u \dzetint} = \dfrac{- g \hfilm^3}{3 \nu} \left[ \left( \dxi \hfilm \right) \cos{\smAng} - \sin{\smAng} \right] \, .
    \label{eq:model-conservativeCurvilinear-q}
\end{eqnarray}
Using the continuity equation integrated over the film thickness, we obtain
\begin{equation}
    \dt \tinySpace \hfilm = - \dfrac{1}{r} \, \dxi \left( qr \right) \label{eq:model-hOfT}
\end{equation}
in conservative form. Equations~\eqref{eq:model-conservativeCurvilinear-q} and~\eqref{eq:model-hOfT} describe the time evolution of the film thickness resulting from the sole gravity-driven drainage, at any location on the stalagmite surface. Drainage is facilitated by either (i) a thicker film, (ii) a film with larger thickness gradients, or (iii) a stronger inclination of the stalagmite wall below the film. However, these last two effects are not additive: on an almost vertical wall ($\smAng \rightarrow \pi / 2 $), the contribution from film thickness variations ($\sim\left( \dxi \hfilm \right) \cos{\smAng}$) becomes negligible compared to the surface inclination term ($\sim \sin{\smAng}$) of Eq.~\eqref{eq:model-conservativeCurvilinear-q}. Hence, for parabolic stalagmites, the local inclination of the surface ($\sim \sin{\smAng}$) drives the drainage for positions that are characterised by $\left( \partial_\xi h \right) \ll \tan{\smAng}$, a limit case to which we refer as inclination-dominated drainage. Conversely, on an almost horizontal stalagmite surface for which $\left( \partial_\xi h \right) \gg \tan{\smAng}$ (also denoted flat stalagmites), drainage will be mostly driven by film thickness gradients ($\sim\left( \dxi \hfilm \right) \cos{\smAng}$).

To connect the local drainage dynamics with the global mass transport, the surface-parallel flux~$\flux$ can be related to the total volumetric drainage flow rate over a circular ring of radius~$r(\xi)$ as $\Qdrain = 2 \pi r(\xi) q$. Assuming near-saturated relative humidity conditions typically encountered in caves (\eg, \cite{GunnEncyclopedia}), evaporation and condensation balance each other out and there should be no other inflow compensating~$\Qdrain$ but the liquid supplied by the successive drops falling into the film. We further hypothesise that these drops do not yield splash at impact to avoid dealing with an additional outflow dependent on the local film thickness (\eg, \cite{Wang2000,Cossali2004}), and that the entire drop volume~$\Vdrop$ is added into the film at each impact. While drop splashing at impact is almost ubiquitous in caves (\eg, \cite{Parmentier2023}), it does not significantly affect the long-term drainage dynamics. 
The discrete inflow of drops, denoted~$\Qdrip$, is given by
\begin{equation}
    \Qdrip = \dfrac{\Vdrop}{\tdrip} \, ,
    \label{eq:discrete-inflow}
\end{equation}
where $\tdrip$ is the dripping period between two successive drops, ranging from less than a second to several weeks in most caves (\eg, \cite{Genty1998,Parmentier2019}). The volume~$\Vdrop$ is approximately constant as it results from the balance between gravitational and surface tension forces when a drop hanging from the ceiling detaches (\eg, \cite{Tateslaw}). Hence, $\Qdrip$ varies solely in response to variable $\tdrip$. We also consider that all the drops land at position $r = 0$,~\ie, that there is no impact point position dispersal. Impact point dispersal can nevertheless be incorporated easily using a 2D description of the problem (\eg, \cite{Parmentier2024} for the case of flat stalagmites), but that lies beyond the present 1D, axisymmetric model. 

A perfect balance between the drop inflow $\Qdrip$ and drainage outflow $\Qdrain$ ensures that the film remains at stationary state. However, because the drops enter the film one at a time, the stationary film thickness should oscillate between a maximum immediately after a new drop has been added to the film, and a minimum right before the next drop addition. In the following, we refer to this minimum film thickness as the stationary film thickness~$\hfilmStnry$, which defines the baseline envelope of the film oscillations. 
Departures from the balance between $\Qdrip$ and $\Qdrain$ should result in transient evolution of the film before it reaches a new stationary state. If the stalagmite surface is initially dry, this transient state corresponds to the propagation of a liquid front on top of the stalagmite, solely induced by the inflow of drops. These phenomena illustrate the complex interplay between localised drop additions in the film and spatially distributed drainage on curved surfaces. In the following, we formalise how the non-linear structure of Eqs.~\eqref{eq:model-conservativeCurvilinear-q} and~\eqref{eq:model-hOfT} leads to various drainage regimes for either flat or parabolic stalagmites in response to variable inflow.

\subsection{Characteristic scales and drainage timescale}\label{sec:charac-scales}

The thin film approximation implies a clear decoupling between characteristic length scales across and along the liquid film (\eg, \cite{Liu2017,Zheng2022}). Typical film thicknesses in caves are of the order of $\SI{100}{\micro\meter}$ while stalagmite radii usually reach a few centimetres. We therefore define two length scales to normalise variables and equations: $\hdrop = \SI{100}{\micro\meter}$ for variations across the film, and $\rdrop = \SI{1}{\centi\meter}$ for variations along the film. These definitions match the typical height and radial extent of the puddle formed by a drop upon impact and correspond to a drop volume of the same order of magnitude ($\Vdrop = \SI{1.6e-8}{\cubic\meter}$ for a paraboloid of radius $\rdrop$ and height $\hdrop$) as in lab experiments ($\Vdrop^{\rm lab} = \SI{5.1e-8}{\cubic\meter}$) and caves ($\Vdrop^{\rm cave} \simeq \SI{8.2e-8}{\cubic\meter}$).

If $\tdrain$ and $\qdrain$ designate the corresponding characteristic drainage timescale and surface-parallel flux scale, respectively, the variables appearing in Eqs.~\eqref{eq:sm-shape}, \eqref{eq:model-conservativeCurvilinear-q} and~\eqref{eq:model-hOfT} now write as
\begin{eqnarray}
    r = r' \rdrop \, \text{,} \; \; \Rsm = \Rsm' \rdrop \, \text{,} \; \;  \xi = \xi' \rdrop \, \text{,} \; \; t = t' \tdrain \, \text{,} \; \; \hfilm = \hfilm' \hdrop \, \text{,} \; \; \eta = \eta' \hdrop \, \text{,} \; \; \shapeFac = \shapeFac' \dfrac{\hdrop}{\rdrop^2} \, \text{,} \; \; \flux = \flux' \qdrain \, \text{,} \nonumber
\end{eqnarray}
where the primes refer to the nondimensional variables. The prime notation will be systematically used in scaling analyses (see Sec.~\ref{subsec:scalings}), but will be dropped for the numerical resolution, where all variables remain nondimensional by construction. 
The nondimensionalisation of the shape factor~$\shapeFac$ preserves the structure of Eq.~\eqref{eq:sm-shape} in nondimensional form. Inserting the above notations into Eqs.~\eqref{eq:model-conservativeCurvilinear-q} and~\eqref{eq:model-hOfT}, it is found that
\begin{eqnarray}
    \partial_{t'} \hfilm' = \dfrac{g \hdrop^3 \tdrain}{\nu \rdrop^2} \dfrac{1}{r'} \;  \dfrac{\partial}{\partial \xi'} \left[ \dfrac{\left(\hfilm'\right)^3}{3} \left( \left( \partial_{\xi'} \hfilm'  \right) \cos{\smAng} - \left(\dfrac{\rdrop}{\hdrop} \right) \sin{\smAng} \right) r' \right] \, . \label{eq:model-nd-scales}
\end{eqnarray}
The drainage timescale is thus defined as
\begin{equation}
    \tdrain = \dfrac{\nu \rdrop^2}{g \hdrop^3} \, , \label{eq:chap3-drainage-time} 
\end{equation}
which is about $\SI{10}{\second}$ for the chosen values of $\rdrop$ and $\hdrop$, taking into account that $g \simeq \SI{10}{\meter\per\square\second}$ and that, for water, $\nu \simeq \SI{e-6}{\square\meter\per\second}$. Correspondingly, the characteristic surface-parallel flux scale is given by
\begin{equation}
    \qdrain = \dfrac{g \hdrop^4}{\nu \rdrop} \, . 
\end{equation}
This scale can also be interpreted as $\qdrain = \rdrop \hdrop / \tdrain$,~\ie, the flux required to drain a volume equivalent to a single drop over a width~$\rdrop$ during one drainage timescale~$\tdrain$. 

In other words, adding drops with a dripping period equivalent to the drainage timescale can serve as an onset to efficiently compare high inflows such that $\tdrip < \tdrain$, and low inflows for which $\tdrip > \tdrain$. However, high inflows such that $\tdrip \ll \SI{1}{\second}$ in dimensional units could, in practice, lead to the emergence of chaotic dripping or transition to jetting (\eg, \cite{Clanet1999}), which is out of the scope of our modelling. The case of a regular jet feeding the film in the limit $\tdrip \rightarrow 0$ will nevertheless be considered later. Finally, the surface inclination term from Eq.~\eqref{eq:model-nd-scales} is observed to be multiplied by the aspect ratio $\rdrop/\hdrop \gg 1$,  reminiscent of the fact that even small surface slopes can dominate over film thickness gradients.

\subsection{Nature of the governing equations and numerical approach}  
\label{subsec:nature-gov-eqs}

    We propose a numerical framework that resolves the time-dependent drainage equations. Using the characteristic scales from the former section, the nondimensional version of Eqs.~\eqref{eq:model-conservativeCurvilinear-q} and~\eqref{eq:model-hOfT} reads
    \begin{eqnarray}
        \flux &=& -\dfrac{\hfilm^{3}}{3} \left[ \left(\dxi\hfilm\right) \cos{\smAng} - \left( \dfrac{\rdrop}{\hdrop} \right) \sin{\smAng} \right] \, , \label{eq:model-q-nd} \\
        \dt \hfilm &=& -\dfrac{1}{r} \dxi \left( \flux r \right) \, , \label{eq:model-h-nd} 
    \end{eqnarray}
    for which no analytical solution exists in the unsteady regime. For the sake of simplicity, variables will now be written without primes but remain nondimensional, as defined in Sec.~\ref{sec:charac-scales}. The stalagmite profile, $\smelv(r)$, is defined according to Eq.~\eqref{eq:sm-shape} for $0 \le r \le \Rsm$, with $\Rsm$ the horizontal truncation of the domain, or the stalagmite radius for flat cases (see Fig.~\ref{fig:chap3-model-bigSchematic}\;(c)). The numerical domain is discretised along the arc length coordinate $\xi(r)$, which defines a one-dimensional computational grid. The local inclination, $\smAng$, and the arc length, $\xi$, are given by
    \[
        \smAng = \arctan{\left(\partial_r \smelv\right)} \, , \quad \xi(r) = \displaystyle{ \int_{0}^{r} \left(1 + \left(\partial_{\tilde{r}} \smelv\right)^{2} \right)^{1/2} \mathrm{d} \tilde{r} } \, ,\label{eq:model-arcLengthDef}
    \]
    respectively. The first cell corresponds to the stalagmite centreline, in $r = 0$, for which $\smAng = 0$.

    The above parabolic system is closed by imposing one initial condition and two boundary conditions. However, it is not possible to apply two Neumann conditions on the flux in both $r = 0$ (symmetry) and $r = \Rsm$ in perfectly flat cases ($\shapeFac = 0$). As detailed in Sec.~\ref{subsec:scalings}, we rather approximate the outer stalagmite wall for $r \ge R$ (see Eq.~\eqref{eq:sm-shape}) by a cone portion of constant opening angle, denoted $\phiStar$, for which the film thickness can be calculated (see Eq.~\eqref{eq:cone-anal}). For inclined stalagmite profiles, the opening angle can be computed from the tangent to the surface at the outer edge of the domain. The cross-sectional flux at the outer edge is then matched with the equivalent flux that would be obtained on a cone with this angle. On the other hand, flat stalagmites require to impose a value for $\phiStar$. 
    Unless specified otherwise, $\phiStar = \SI{45}{\degree}$ is chosen as representative of flat stalagmites (\eg, \cite{Szymczak2025}). 
    
    The film is initialised as dry everywhere, \ie, $\hfilm(r, t=0) = 0$ $\forall r$, until drops are added. At early times, when the film front advances over the solid surface, only the upstream boundary condition in $r = 0$ remains relevant due to the locally hyperbolic nature of the system. 
    By axisymmetry, drainage at the stalagmite centreline should only occur in the downstream direction ($r > 0$). 
    This translates into a homogeneous Dirichlet condition for the flux~$\flux$, and a homogeneous Neumann condition for the film thickness~$\hfilm$ at the outer face of the first cell of the numerical spatial domain. 

     The equations are solved based on a a conservative discretisation formalism. A standard 4th-order explicit Runge-Kutta scheme is used for time integration with a fixed global time step, which is reduced as needed to maintain stability. Typically, $\Dt = \SI{e-4}{}$ and $\Dxi = \SI{0.05}{}$ to limit the computational cost of long time integration cases.  Comparison with a much more refined case shows that the error is lower than $\SI{3}{\percent}$ for the film thickness at $r = 0$.

\begin{figure}
\centerline{\includegraphics[width = \linewidth]{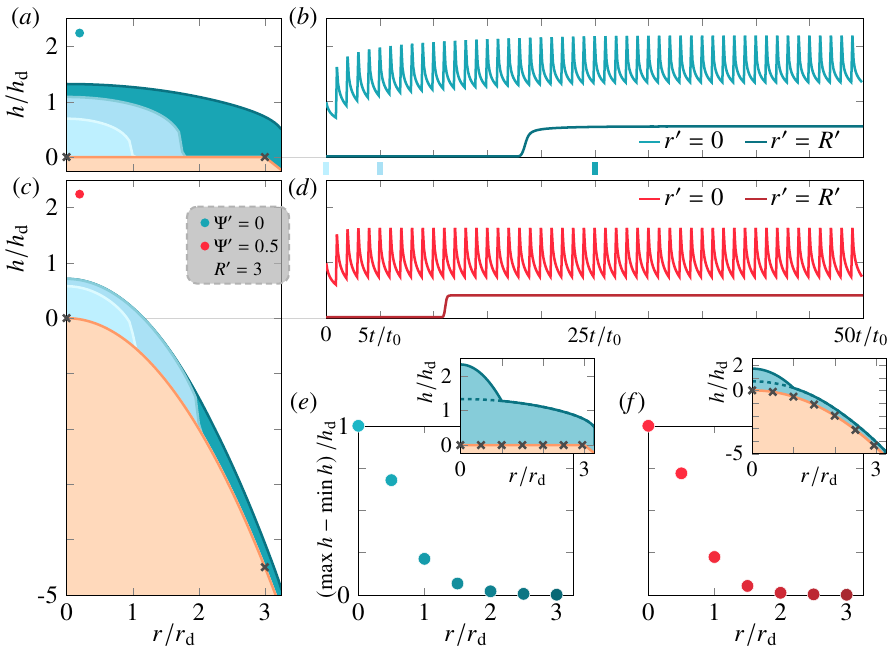}}
    \caption{Numerical film thickness evolution in space (a, c) and time (b, d), for flat ($\shapeFac' = 0$) and slightly curved parabolic ($\shapeFac' = 0.5$) stalagmites, for $\tdrip' = 1$. Panels (a) and (c) show the evolution of the film thickness (in blue) over the stalagmite (in beige), after 1, 5 and 25 drops were added and have spread into the film. Panels (b) and (d) show the corresponding film thickness evolution with the successive drop impacts at the centre (in $r' = 0$), and very close to the edge of the stalagmite. These radial positions are indicated by black crosses in (a) and (c). The film reaches a stationary state after about 25 drops in the flat case and 3 drops in the inclined case. Panels (e) and (f) represent the difference between the two envelopes of the oscillating signals, \ie, the difference between the maximum and minimum film thickness values at stationary state shown in (b) and (d) for various radial positions, illustrated in the corresponding inset by the black crosses. The radius/domain size was set to $\Rsm' = 3$ for both geometries. The profile $\hfilmStnry'(r')$ just before an impact $\left( t' = k\tdrip'^{-}, k \in \mathbb{N} \right)$ is shown in (a), (c), and insets of (e), (f) in dark blue. The legend in the grey box refers to all panels. Primes denote nondimensional variables where explicit normalisation is absent. See also Supplementary Movies 1 and 2 for similar cases ($\tdrip = 1$, $\Rsm' = 5$ and $\shapeFac' = 0$ and 1, resp.). 
    } 
    \label{fig:flat-vs-incl-example}
\end{figure}

    \subsection{Discrete drop inflow}\label{sec:chap3-dropAdd}

    An inflow of successive drop impacts is added over the stalagmite to create the liquid film and maintain it at stationary state. To include this discrete inflow of drops, Eq.~\eqref{eq:model-hOfT} is reformulated by including punctual source terms as follows: 
    \begin{equation}
        \dt \hfilm + \dfrac{1}{r} \miniSpace \dxi\left( q r \right) = \sum_k{ \hdroplets \tinySpace (\xi \microSpace) \, \delta\hspace{-2.5pt}\left(t - k \tdrip \right) } \, \text{,} \; k \in \mathbb{N} \, , \label{eq:source-terms}
    \end{equation}
    where $\delta$ is the Dirac distribution and $\hdroplets(\xi)$ is the drop-shaped film thickness increment. To approach the early-stage shape of a drop crushing at impact (\eg, \cite{Rioboo2002}), we model each drop addition in the film using a parabolic profile spread over a radius~$r = 1$ and height~$h = 1$. 
    The corresponding dimensional drop volume, $\pi \rdrop^2 \hdrop / 2$, sets the characteristic scale for the nondimensionalisation. The nondimensional drop volume is thus equal to 1 in the simulations, and the drop inflow is solely set by the dripping period $\tdrip$. The exact added drop shape has limited effect on subsequent film thickness variations. This approach does not capture the exact dynamics of drop impact since inertial and capillary effects were not taken into account in Eqs.~\eqref{eq:model-conservativeCurvilinear-q} and~\eqref{eq:model-hOfT}. However, the relevant timescales remain consistent with experimental measurements as drops can be viewed as instantaneously added to the film (a drop impact in a thin film typically lasts for $\lesssim \SI{200}{\milli\second}$ as shown in, \eg, \cite{Wang2004,Parmentier2023}).

\section{Numerical results} 
\label{subsec:num-examples}

    Using the numerical resolution described in the former section, we illustrate how the interplay between discrete drop inflow and geometry-constrained drainage outflow modulates the stationary state of the film. We first describe the general phenomenology, then explore the response of the film stationary state to $r, t, \tdrip$ and $\shapeFac$ (or $\Rsm$ in the flat case). We then briefly describe the transient spreading of the film over the stalagmite before stationarity is achieved, denoted as the propagating front or the propagating tip of the film. 
    All quantities in this section are nondimensional and written without primes. Legends and captions clarify the use of dimensional variables where needed, for consistency with the rest of the paper. Finally, 
    unless stated otherwise and when available, numerical fit exponents were chosen to match theoretical predictions which will be developed later in Sec.~\ref{subsec:scalings}. Slight deviations in their values should not significantly affect their interpretation. 
    
    \subsection{Phenomenology}

    We first examine the general phenomenology of the film spreading over an initially dry stalagmite of variable shape and/or with variable drop inflow.  Figure~\ref{fig:flat-vs-incl-example} displays the film response to a fixed drop inflow such that $\tdrip = 1$ (\ie, intermediate inflow) over two stalagmite profiles with identical domain size ($\Rsm = 3$): $\shapeFac = 0$ (a, b, e), $\shapeFac = 0.5$ (c, d, f). Panels (a) and (c) show snapshots of the film shape over the surface. Panels (b) and (d) track the corresponding oscillating film thickness at the centre ($r = 0$), \ie, where the drop impacts the stalagmite, and at the domain outer limit ($r \simeq \Rsm$). Supplementary Movies 1 and 2 further show the time evolution of the film for two similar cases ($\tdrip = 1$, $\Rsm = 5$, $\shapeFac = 0$ and 1, respectively).
    
    In Figs.~\ref{fig:flat-vs-incl-example}\;(a-d), the film thickness at the centre gradually increases with the successive drop impacts, progressively extending the film coverage over the entire stalagmite surface considered in the simulations. The resulting film is thicker for the flat stalagmite than for the parabolic one.
    The film eventually reaches a stationary state, but more rapidly in the parabolic case (panel (d)). 
    The number of drop impacts after which the difference between two successive peaks falls below, \eg, \SI{1}{\percent} ($k$ such that $\left|\hfilm\left(t - k \tdrip\right) - \hfilm\left(t - (k-1) \tdrip\right)\right|/\hfilm\left(t - k \tdrip\right) \le 0.01$ in $r = 0$) is indeed 6 in the parabolic case, against 24 in the flat case. 
    Similarly, although the film propagates over a longer distance in the parabolic case for a given number of drops, it reaches the outer limit of the domain after fewer drop impacts (11 in the parabolic case, against 18 in the flat case). This is due to the faster and more efficient drainage on inclined geometries, which reduces the local liquid accumulation favoured in flat cases. Consequently, the stationary state at $r = 0$ is also attained once the film has completely covered the stalagmite in the flat case (panels (a, b)), while partial coverage is sufficient in the parabolic case (panels (c, d)).

    In Figs.~\ref{fig:flat-vs-incl-example}\;(e) and (f), the oscillation amplitude (\ie, the difference between the local maximum and minimum film thickness between two successive drops) is reported as a function of the distance to the impact point for both geometries, at stationary state. Variations are substantial for $\xi \lesssim 1$ in panels (e, f), but beyond this region corresponding to the drop radius, the oscillations are quickly damped and the film thickness becomes time-independent. Interestingly, while even a slightly curved parabolic stalagmite ($\shapeFac < 1$) affects the overall film thickness response (panels (c, d)), the stalagmite shape has little influence on the damping of these oscillations if it is not sufficiently inclined.

\begin{figure}
\centerline{\includegraphics[width = \linewidth]{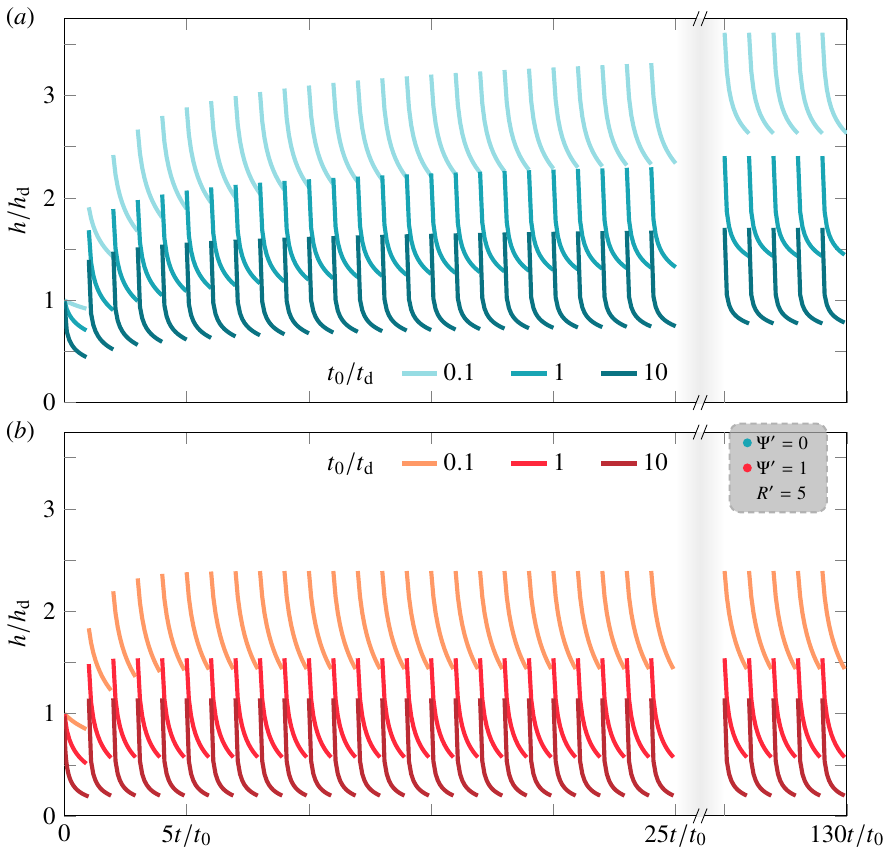}}
    \caption{Numerical film thickness evolution in time for different inflow frequencies over (a) flat ($\shapeFac' = 0$) and (b) inclined geometries ($\shapeFac' = 1$), for a domain size $\Rsm' = 5$. The dripping period was varied between high ($\tdrip' = 0.1$) and low inflows ($\tdrip' = 10$). Both graphs show the first 25 drop additions in the film, then from the 125$^{\text{th}}$ to the 130$^{\text{th}}$ drop addition with a 100-drop long gap indicated by the grey shaded area. The film over the flat stalagmite (a) only reaches a stationary state beyond the first 25 drops shown in the graphs, after about 96 drops ($\tdrip' = 0.1$), 54 drops ($\tdrip' = 1$) and 31 drops ($\tdrip' = 10$). The legend in the grey box refers to all panels. Primes denote nondimensional variables where explicit normalisation is absent. 
    }
    \label{fig:inflow-variable}
\end{figure}

    Figure~\ref{fig:inflow-variable} shows the temporal response to different dripping frequencies of the film thickness at the centreline ($r = 0$) of a flat stalagmite (panel (a)) and of a parabolic stalagmite with $\shapeFac = 1$ (panel (b)) over a larger radius/domain size $\Rsm = 5$. Three inflow regimes are considered, as defined in Sec.~\ref{sec:charac-scales}: high ($\tdrip < 1$), intermediate ($\tdrip = 1$) and low ($\tdrip > 1$). Comparing the intermediate-inflow cases from Figs.~\ref{fig:flat-vs-incl-example}\;(d) and \ref{fig:inflow-variable}\;(b), the film reaches a stationary state after a similar number of drops in the parabolic case. Over a flat stalagmite though, this stationary state is delayed as it is only reached after 54 drop impacts for $\Rsm = 5$ (Fig.~\ref{fig:inflow-variable}\;(a)), by contrast with 24 drop impacts for $\Rsm = 3$ (Fig.~\ref{fig:flat-vs-incl-example}\;(b)). Furthermore, in Fig.~\ref{fig:inflow-variable}, it can be observed that increasing the inflow leads to a larger film thickness for both geometries, as expected from Eq.~\eqref{eq:model-conservativeCurvilinear-q}. 
    Increasing the inflow at fixed $\tdrain$ also delays the appearance of the stationary state, as it can be most clearly seen in Fig.~\ref{fig:inflow-variable}\;(a) for the high inflow over the flat case, where stationarity is reached after about 96 drop impacts. 
    By contrast, as illustrated in Fig.~\ref{fig:inflow-variable}\;(b), this delay is not as pronounced for a parabolic stalagmite because of the faster drainage in this configuration. In dimensional terms though, the viscous relaxation time would remain comparable across regimes since the stationary state depends on viscous smoothing.

\subsection{Stationary regime}
\label{sec:stationary}

The stationary state of the film provides a physically meaningful and reproducible indicator of the system. We now explore in further details its response to variable inflow and stalagmite geometry, first through numerical measurements of the film thickness. Experimental measurements will be compared in Sec.~\ref{sec:exp}. In all the subsequent numerical and experimental measurements, the stationary film thickness, denoted as $\hfilmStnry$, is consistently defined as the minimum value reached between two successive drop impacts once the transient regime of film filling is over: 
\[
    \hfilmStnry(r) \equiv \min_t{\left( \hfilm(r, t) \right)} \, .
\]
This definition captures the long-term evolution of the film dynamics while excluding transient effects associated with individual drop impacts. It also ensures a robust comparison between numerical simulations and experiments.  
The stationary film thickness is denoted $\hfilmStnry(r)$, and its value at the stalagmite centreline is $\hfilmStnryZ = \hfilmStnry(r=0)$. For all cases, we consider that the film is at stationary state once the film tip has reached the outer limit of the domain (typically after $\sim100$ drop impacts). For completeness, the limiting case of pure gravity-driven drainage without drop inflow is covered in \cite{Parmentier2024}.

\begin{figure}
\centerline{\includegraphics[width = \linewidth]{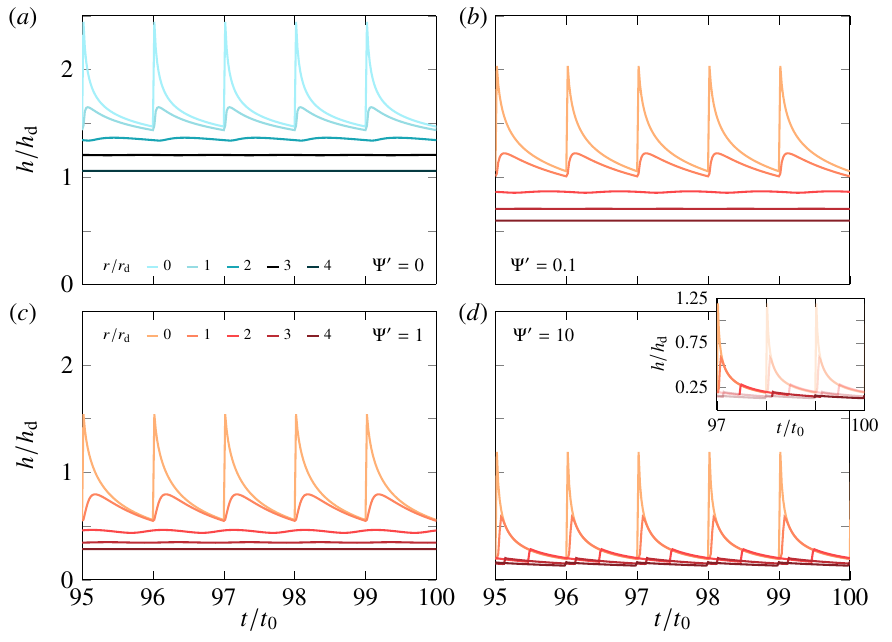}}
    \caption{Numerical film thickness over time for $\tdrip' = 1$, at various radial positions $r' \in \left\{ 0, 1, 2, 3, 4 \right\}$, for $\Rsm' = 5$ and (a) $\shapeFac' = 0$, (b) $\shapeFac' = 0.1$, (c) $\shapeFac' = 1$ and (d) $\shapeFac' = 10$. Each panel presents 5 successive drop impacts at stationary state. The inset of panel (d) shows the maximum film thickness reached at the different positions but corresponding to a single drop impact. 
    It highlights how the successive perturbations of individual drops become progressively shifted along the surface. Primes denote nondimensional variables where explicit normalisation is absent.
    }
    \label{fig:hs_r}
\end{figure}

    In Fig.~\ref{fig:hs_r}, the thickness of the film at stationary state is shown as a function of time at several radial positions ($r={0,1,2,3,4,5}$) for $\Rsm=5$ and shapes ranging from flat ($\shapeFac=0$, panel (a)) to increasingly curved profiles ($\shapeFac=0.1,1,10$, panels (b–d)). The film thickness increases and decreases in response to drop additions. However, the local maximum reached by the film thickness over a given dripping period is progressively delayed with increasing distance from the drop impact point. In accordance with Figs.~\ref{fig:flat-vs-incl-example}\;(e-f), this perturbation becomes negligible beyond a certain radius, and the thickness profile becomes time-independent. In Fig.~\ref{fig:hs_r}\;(d) representing the most curved parabolic profile, a pronounced maximum of film thickness is seen at every sampled position.
    The maximum can even occur after the next drop has already been added to the film (also slightly visible in panel (c)). This behaviour is emphasised by the inset of Fig.~\ref{fig:hs_r}\;(d) where the darker colours show the maxima related to the same drop. 
        
    \begin{figure}
    \centerline{\includegraphics[width = \linewidth]{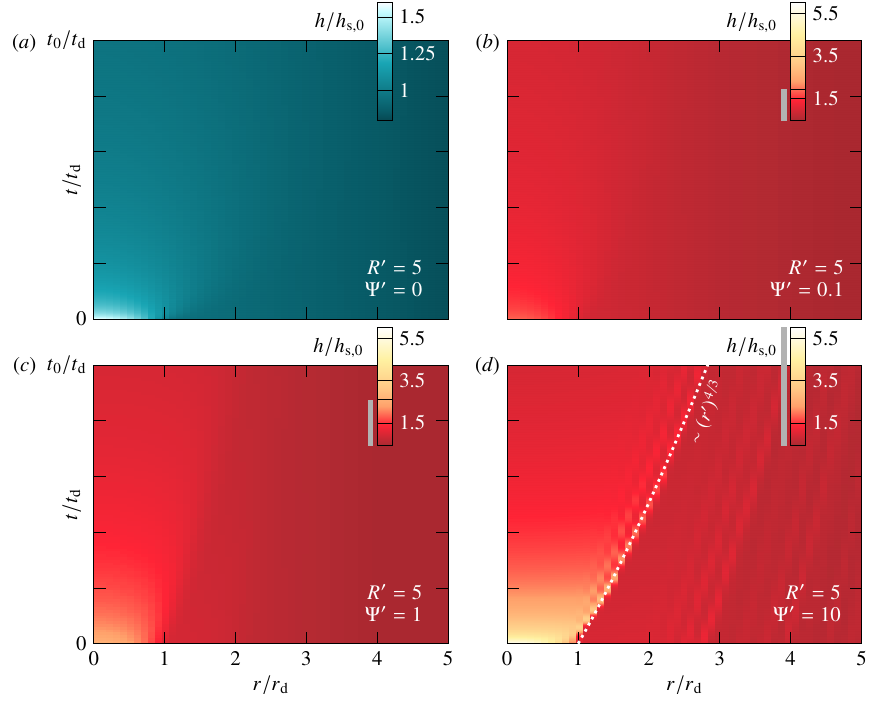}}
        \caption{Spatio-temporal evolution of the numerical film thickness at stationary state,~$\hfilm/\hfilmStnryZ$, in between two successive impacts, for $\tdrip' = 1$, $\Rsm' = 5$ and four shape factors: (a) $\shapeFac' = 0$, (b) $\shapeFac' = 0.1$, (c) $\shapeFac' = 1$ and (d) $\shapeFac' = 10$. The colour bar from (a) corresponds to panel (a). The colour bar from (d) is shared among (b) and (c), but it is accompanied with grey level bars showing the portion of the colour bar actually covered by each heat map. Note that all colour bars start at $\sim 0.6$ (\ie, $\hfilm/\hfilmStnryZ < 1$). The dashed line from (d) shows that the front can be estimated as evolving along $\sim \left(r' \right)^{4/3}$. Primes denote nondimensional variables where explicit normalisation is absent. 
        }
        \label{fig:heatMap}
    \end{figure}

    \begin{figure}
    \centerline{\includegraphics[width = \linewidth]{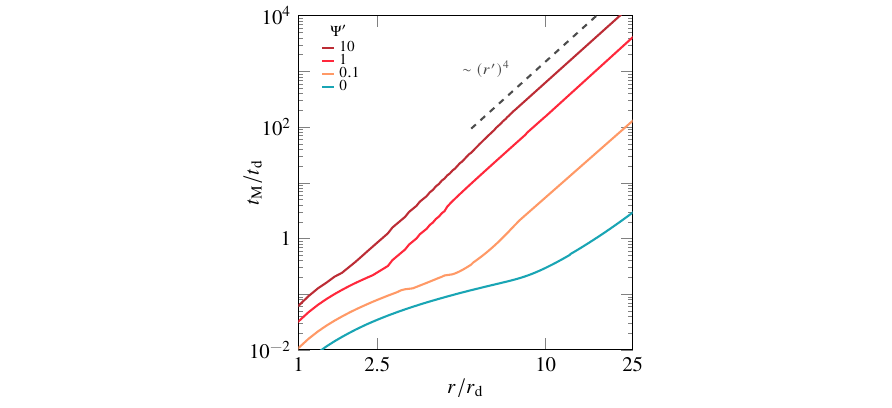}}
        \caption{Time at which the maximum film thickness is reached after a drop impact as a function of the curvilinear coordinate for $\tdrip' = 1$, $\Rsm' = 5$ and four shape factors: $\shapeFac' = 0$, $\shapeFac' = 0.1$, $\shapeFac' = 1$ and $\shapeFac' = 10$. The dashed lines represent the scaling obtained for the parabolic case ($\tMax' \sim \left(r'\right)^4$). Primes denote nondimensional variables where explicit normalisation is absent.
        } 
        \label{fig:noone}
    \end{figure}

    The effect of the drop addition in the film can also be seen in Fig.~\ref{fig:heatMap}, which presents complementary heat maps of the film thickness as a function of time and position. As in Fig.~\ref{fig:hs_r}, panels (a-d) of Fig.~\ref{fig:heatMap} correspond to a flat case ($\shapeFac = 0$, panel (a)) and parabolic cases for which $\shapeFac = 0.1$, 1 and 10 in panels (b), (c) and (d), respectively. 
    The heat maps show the film thickness at stationary state during one dripping period, following a drop addition. In this particular figure, the film thickness is normalised by its stationary value at the centreline ($\hfilm(r,t)/\hfilmStnryZ$), and not by $\hdrop$ as in Fig.~\ref{fig:hs_r}, to enhance the relative contrast in the spatio-temporal response of the film within one dripping period. In all heat maps, the film appears the thickest at early times ($t \rightarrow 0$) close to where the drop was added in the film, \ie, over a radius $\xi \simeq 1$. 
    The perturbation caused by the individual drop away from this region is progressively damped over time and position in panels (a, b), in line with the observations from Fig.~\ref{fig:hs_r}. In contrast, in Fig.~\ref{fig:heatMap}\;(d) a series of sharp ripple-like fronts are observed for the steepest parabolic stalagmite ($\shapeFac = 10$). Only one sharp front is visible in panel (c). The faint oscillations visible for $\xi \gtrsim 2$ in Fig.~\ref{fig:heatMap}\;(d) correspond to residual perturbations induced by previous drop impacts. 
    These delayed perturbations reflect a wave-like advection of the successive drops, consistent with the overlapping patterns highlighted in the inset of Fig.~\ref{fig:hs_r}. The front can be described as evolving according to $\xi^{4/3}$. 
    Although in all the previous figures, the film appeared the thinnest in absolute value for more curved stalagmites, the relative variations with respect to the stationary film thickness increase with increasing $\shapeFac$ in Fig.~\ref{fig:heatMap}.

    From the combined observations of Figs.~\ref{fig:flat-vs-incl-example}, \ref{fig:hs_r} and \ref{fig:heatMap}, we first infer the existence of a characteristic scale, denoted $\rStnry$, beyond which the discrete nature of the drop inflow becomes negligible. 
    This scale separates an unsteady region, strongly perturbed by discrete drop impacts, from a steady region where the discrete nature of the inflow becomes negligible and the film behaves as if continuously fed. In the steady region, the film shape converges toward a stationary profile consistent with the profile expected under continuous inflow conditions. This stationary profile will be derived analytically (see Sec.~\ref{subsec:scalings}). The dominant attenuation mechanism of the drop-induced perturbations beyond this characteristic length scale also seems dependent on the stalagmite shape. The heat maps of Fig.~\ref{fig:heatMap} reveal a transition from localised, diffusive-like spreading in the flat and slightly curved cases (panels (a, b)) to advective-like front propagation in parabolic stalagmites involving steeper surfaces (panel (d)). The dynamics in $\shapeFac = 1$ appears intermediate between the diffusive and advective regimes of the perturbation propagation. Parabolic configurations thus promote perturbation transport over longer distances and delayed times, even when the film is at stationary state.
    
    Figure~\ref{fig:noone} further characterises the perturbation propagation highlighted above. It illustrates how the time at which the maximum film thickness is reached, $\tMax$, evolves with the position. This time is defined as 
    \begin{equation}
        \tMax(r) \equiv \arg{\,\max_{t}{\,\hfilm(r, t)}} \, , \quad k\tdrip \le t \le (k+1)\tdrip, \quad k \in \mathbb{N} \, . 
    \end{equation}
    over one dripping period when the film is at stationary state. The time of maximum thickness increases with the position and for increasing shape factor $\shapeFac$, in accordance with the observed delayed propagation of the perturbation. 
    
    Although the perturbation front advects faster on more curved geometries, the local time of maximum thickness $t_M$ increases with $\shapeFac$, as the stronger drainage delays the moment when the film locally reaches its maximum. Consequently, wave superposition appears and the maximum reached at a position away from the centre at a given time may appear shifted, as seen in the overlapping wave patterns from the inset of Fig.~\ref{fig:hs_r}\;(d), and Fig.~\ref{fig:heatMap}\;(d).   
    In the near-field at short distances, the time of maximum thickness reflects the post-impact adjustment of the film thickness to the drop perturbation. In particular, $\tMax$ shows a distinct behaviour in the near-field for $\shapeFac \lesssim 1$, \ie, for shapes over which diffusion was observed to damp the perturbations associated with the drop impacts. In the far-field at large radii, it scales with the position $r$ for parabolic geometries as $\tMax \sim r^{4}$. 
    From there, if we consider that the unsteady region corresponds to the position at which $\tMax(\rStnry) = \tdrip$, this scaling becomes 
    $\rStnry \sim \tdrip^{1/4}$. 
    This would mark the transition where successive drop responses start to overlap, separating time-dependent and time-independent areas. 
    In the flat case, the interpretation of the time $\tMax$ becomes less relevant in the far-field, as perturbations are essentially diffused in the near-field and the film thickness no longer varies significantly over time. 
    
    \begin{figure}
    \centerline{\includegraphics[width = \linewidth]{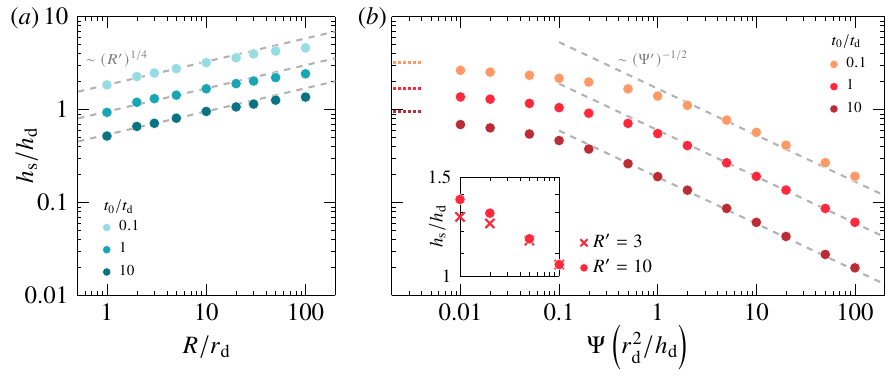}}
        \caption{Numerical stationary film thickness at the centreline, $\hfilmStnryZ/\hdrop$, for different geometry and drop inflow: (a) in the flat case, as a function of the stalagmite radius~$\Rsm$, (b) in the inclined case, as a function of the shape factor~$\shapeFac$ for a domain size $\Rsm = 10$. The grey dashed lines in both graphs show the respective scalings obtained, \ie, $\hfilmStnryZ' \sim \Rsm'^{1/4}$ in (a), and $\hfilmStnry \sim \shapeFac^{-1/2}$ in (b). All the series shown in (b) depart from the power law for $\shapeFac' \lesssim 1$ and tend to the flat case for $\Rsm' = 10$, with respect to their inflow. The dotted lines for $\shapeFac' \ll 1$ represent these asymptotic regimes for each inflow. The inset in (b) illustrates the variations of $\hfilmStnry$ with $\shapeFac'$ for $\shapeFac' \ll 1$, $\tdrip' = 1$, and two different radii $\Rsm' = 3$ and 10. The abscissa range from the inset is identical to the main graph. Primes denote nondimensional variables where explicit normalisation is absent. 
        } 
        \label{fig:hS_num}
    \end{figure}
    
    Beyond the spatio-temporal propagation of the perturbation, the stationary film thickness, $\hfilmStnry$, also evolves with both the inflow and stalagmite shape. 
    Panel (a) of Fig.~\ref{fig:hS_num} shows $\hfilmStnryZ$ at the centreline for various stalagmite radii~$\Rsm$ in a flat case, for the same three inflow regimes as in Fig.~\ref{fig:inflow-variable}: high ($\tdrip < 1$), intermediate ($\tdrip = 1$) and low ($\tdrip > 1$). Similarly, panel (b) represents $\hfilmStnryZ$ as a function of the shape factor~$\shapeFac$ for parabolic geometries, covering the same three inflow regimes as panel (a). 
    As seen in Fig.~\ref{fig:hS_num}\;(a), $\hfilmStnryZ$ increases with the stalagmite radius~$\Rsm$, in agreement with the accompanying increasing liquid accumulation in this case. For the same reason, $\hfilmStnryZ$ is observed to increase with increasing drop inflow, \ie, with decreasing dripping period~$\tdrip$. From rounded fits performed on our numerical measurements, it is found that the stationary film thickness actually follows 
    \[
        \left. \hfilmStnryZ \, \vphantom{\dfrac{1}{1}} \right|_{\rm flat} \sim \Rsm^{1/4} \tdrip^{-1/4} \, .
    \]
    This behaviour is represented by dashed lines in Fig.~\ref{fig:hS_num}\;(a). A small discrepancy between the fit and the numerical values emerges at large radii ($\Rsm \gg 10$), which may reflect second‑order effects not captured by the above power law.

    Conversely, Figure~\ref{fig:hS_num}\;(b) shows that the stationary film thickness decreases with increasing shape factor~$\shapeFac$, for $\shapeFac \gtrsim 1$. Steeper stalagmite walls should effectively improve the efficiency of the gravity-induced drainage, thereby easing liquid depletion at the centre of the stalagmite. In agreement with flat cases, the stationary film thickness also increases for increasing inflow. The following scaling law fits the numerical simulations well: 
    \[
        \left. \hfilmStnryZ \, \vphantom{\dfrac{1}{1}} \right|_{\rm parab.} \sim \shapeFac^{-1/2} \tdrip^{-1/2} \, , \quad \shapeFac \gtrsim 1 \, .
    \]
    A departure from this law is observed for $\shapeFac \ll 1$, \ie, for stalagmites with small curvature. The stationary film thickness asymptotically tends toward the value obtained for a perfectly flat stalagmite of similar radius as the domain size used. The inset of Fig.~\ref{fig:hS_num}\;(b) shows that $\hfilmStnry$ is slightly influenced by the stalagmite $\Rsm$ for $\shapeFac \ll 1$. 
    
\begin{figure}
    \centerline{\includegraphics[width = \linewidth]{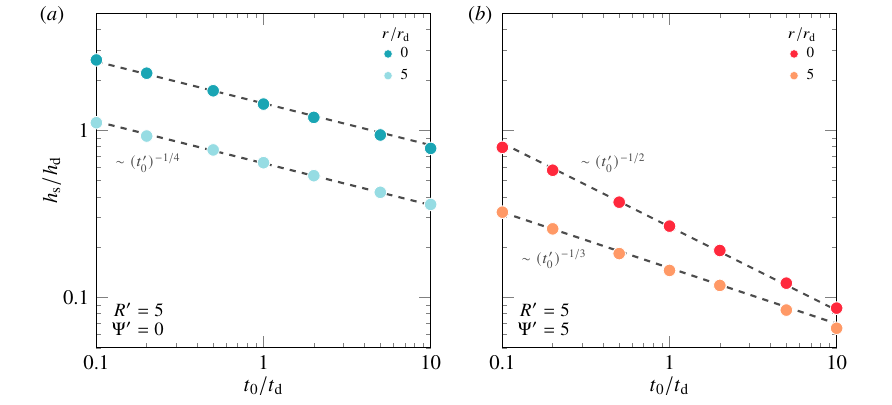}}
        \caption{Evolution of the numerical stationary film thickness, $\hfilmStnry/\hdrop$, with the dripping period, for $\Rsm' = 5$ and different geometries: (a) $\shapeFac' = 0$, and (b) $\shapeFac' = 5$. The dashed lines represent the scalings obtained from curve fitting in Fig.~\ref{fig:hS_num}, in $r = 0$, ($\sim (\tdrip')^{-1/4}$ in (a) and $\sim (\tdrip')^{-1/2}$ in (b)), and at the domain outer boundary, in $r \rightarrow \Rsm$, ($\sim (\tdrip')^{-1/4}$ in (a) and $\sim (\tdrip')^{-1/3}$ in (b)). Primes denote nondimensional variables where explicit normalisation is absent.} 
        \label{fig:nooneb}
    \end{figure}

    The scaling obtained for the evolution of the stationary film thickness with the dripping period, $\hfilmStnryZ \sim \tdrip^{-1/2}$, was evaluated at the centreline. In Fig.~\ref{fig:nooneb}, the stationary film thickness is also reported as a function of the dripping period for positions $r = 0$ and $r = 5$, away from the centreline. Panel (a) shows that, in the flat case, the scaling remains unaffected by the radial position. In contrast, panel (b) illustrates the parabolic case for a representative curvature of $\shapeFac = 5$, where a distinct scaling law emerges:
    \[
        \left. \hfilmStnry \, \vphantom{\dfrac{1}{1}} \right|_{\rm parab.} \sim \tdrip^{-1/3} \, .
    \]   
    Hence, the behaviour of the film at stationary state differs away from the centreline for parabolic stalagmites. This sheds light on a global effect of the geometry on drainage in the parabolic case. By contrast, in the flat case, the scaling obtained did not vary with the radial position.

\subsection{Transient front propagation} \label{sec:front}

Heretofore, we have primarily examined the stationary behaviour of the film in response to a discrete inflow. We focus in this section on the early transient dynamics of the film propagation over the stalagmite, from when the first drop is added. The film advances over the stalagmite surface by covering an increasing area over time. The position of the propagating film tip, denoted by $\lFront$, marks the outermost limit of the wetted area.
We consider two asymptotic limits of the discrete drop inflow: (i) $\tdrip \rightarrow \infty$ which corresponds to a single drop of constant volume, and (ii) $\tdrip \rightarrow 0$, \ie, a continuous inflow. Numerically, while imposing a constant volume simply corresponds to adding one drop on the stalagmite top surface, inner boundary condition should be modified for the continuous inflow. This jet inflow is considered laminar and is approximated by using a non-zero Dirichlet condition on the flux $\flux$ at the centreline using $\Qdrip$. The exact derivation of this condition is reviewed in App.~\ref{app:chap3-bc-num}.

    \begin{figure}
    \centerline{\includegraphics[width = \linewidth]{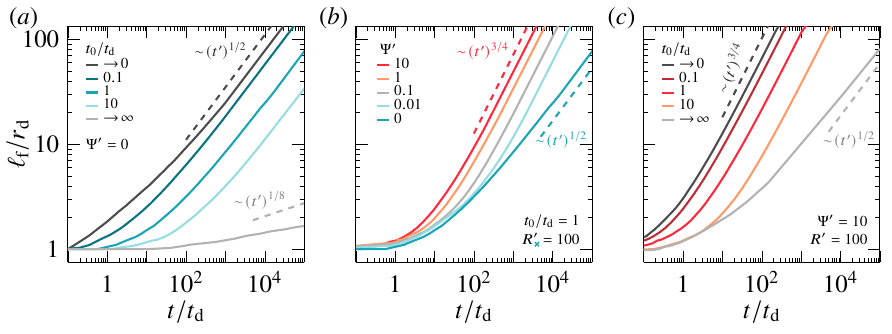}}
        \caption{Numerical front propagation over the stalagmite as a function of time: (a) in a flat case for variable inflow ($\shapeFac' = 0$), (b) for fixed inflow and different stalagmite shapes (from $\shapeFac' = 0$ to 10), and (c) in a parabolic case for variable inflow ($\shapeFac' = 10$). The inflow is varied from a continuous inflow ($\tdrip' \rightarrow 0$) to a constant volume ($\tdrip' \rightarrow \infty$), with in-between discrete drop inflows $\tdrip'$ set to 0.1 (high inflow), 1 (intermediate) and 10 (low). The dashed lines represent the scalings estimated numerically (b, c). The length $\lFront'$ corresponds to a radial position in (a). The blue line from (a) and red line from (c) find their correspondence in (b). 
        Primes denote nondimensional variables where explicit normalisation is absent.  
        }
        \label{fig:front-propagation}
    \end{figure}

We solve the three cases of constant volume spreading and continuous and discrete inflows for different geometries. Figure~\ref{fig:front-propagation} shows the propagating front position, $\lFront$, in each of these regimes for variable stalagmite shapes. 
The cases covered in panel (a) for a flat surface correspond to the constant drop volume ($\tdrip \rightarrow \infty$), low ($\tdrip = 10$), intermediate ($\tdrip = 1$) and high inflows ($\tdrip = 0.1$) and, finally, continuous inflow ($\tdrip \rightarrow 0$). In panel (c), the inflow is varied similarly but for a prescribed stalagmite profile of $\shapeFac = 10$, \ie, a strong curvature. Panel (b) bridges the two other ones as it explores the effect of stalagmite curvature through variations in the shape factor $\shapeFac$ at fixed drop inflow $\tdrip = 1$, going from a flat case ($\shapeFac = 0$) through small values $\shapeFac \ll 1$ and up to $\shapeFac = 10$. 

In general, the progression of $\lFront$ gets faster with increasing inflow. In Fig.~\ref{fig:front-propagation}\;(a), numerical fits on the curves yield 
\[
    \left. \xiFront \vphantom{\dfrac{a}{a}} \right|_{\rm flat} \sim \left\{
    \begin{array}{lr}
        t^{1/8} \, \text{,} &\qquad \text{(constant volume)}\, ,  \\
        t^{1/2} \, \text{,} &\qquad \text{(continuous inflow)}\, , 
    \end{array}
    \right. \quad \shapeFac = 0 \, .
\]
The discrete drop inflows all follow the same scaling as the continuous inflow case at large times, \ie, they scale as $\sim t^{1/2}$. However, at early times, the curves associated with the discrete drop inflows seem to blend with the constant volume case. Each curve can be seen departing from this constant volume baseline around a time corresponding to their relative dripping period, the most visible occurrence being for $\tdrip = 10$. Hence, only a second drop addition in the film seems sufficient to yield a different regime than the constant volume one. 
The crossover between the initial constant volume regime and the asymptotic continuous inflow becomes larger with decreasing inflow, \ie, with increasing dripping period, up to a constant rescaling factor. The same conclusions can be drawn from panel (c) for parabolic geometries. However, the two limiting scalings are different. It is found that the front position evolves as
\[
    \left. \xiFront \vphantom{\dfrac{a}{a}} \right|_{\rm parab.} \sim \left\{
    \begin{array}{lr}
        t^{1/2} \, \text{,} &\qquad \text{(constant volume)}\, ,  \\
        t^{3/4} \, \text{,} &\qquad \text{(continuous inflow)}\, , 
    \end{array}
    \right. \quad \shapeFac > 1 \, .
\]
from numerical fits on the curves of Fig.~\ref{fig:front-propagation}. The front position thus propagates faster over more curved profiles, as expected. Changing the overall shape of the stalagmite also modifies the response of the front propagation at fixed inflow. In Fig.~\ref{fig:front-propagation}\;(b), the flat case ($\shapeFac = 0$) sets the initial behaviour of the front. All the curves at finite $\shapeFac$ values follow this initial trend before departing from it earlier for larger $\shapeFac$, \ie, for more curved profiles. Finally, there is a similarity between the scalings obtained for a front propagating over a flat stalagmite at constant (or discrete) inflow, and for a constant initial volume over an inclined stalagmite. 
The corresponding analytical limits of the regimes described in this section are discussed in Sec.~\ref{subsec:scalings}, where we come back to the dimensional form of the drainage equations.

\section{Scaling analysis}
\label{subsec:scalings}

In this section, we perform a theoretical analysis of the dimensional form of the drainage equations, corresponding to Eqs.~\eqref{eq:model-conservativeCurvilinear-q} and \eqref{eq:model-hOfT}. First, a time-independent solution to the film thickness is sought, for both flat and parabolic stalagmites. An analysis of time-dependent perturbations of the stationary film thickness profile is then made. Finally, an unsteady solution is developed for the parabolic case in the vicinity of the stalagmite centreline. The scaling laws obtained throughout this analysis are compared to the numerical results of Sec.~\ref{subsec:num-examples}. All dimensional variables will be written without primes, while the nondimensional variables will be written using the prime notation introduced in Sec.~\ref{sec:charac-scales}.

\subsection{Stationary film thickness} 

In the former section, Figs.~\ref{fig:flat-vs-incl-example} and \ref{fig:hs_r} revealed that, sufficiently far from the centreline (\ie, for $r \gg 1$ or $\xi \gg 1$), the film thickness becomes time-independent, namely, it is no longer influenced by the discreteness of the drop inflow. The film thickness thus corresponds to the stationary film thickness $\hfilmStnry$ introduced in Sec.~\ref{subsec:num-examples}. Additionally, $\dt \hfilmStnry = 0$ and the drainage equation becomes
\begin{equation}  \label{eq:TimeIndep}
    \flux = \frac{g \hfilmStnry^3}{3 \nu} \left( \sin \smAng - \left( \dxi \hfilmStnry \right) \cos \smAng \right) = \frac{\Vdrop}{2 \pi r \tdrip} \, .
\end{equation}
While there is no closed-form solution to this differential equation, asymptotic regimes may be obtained when either term in the brackets dominates the other. 

\subsubsection{Flat stalagmite}

\begin{figure}
    \centerline{\includegraphics[width = \linewidth]{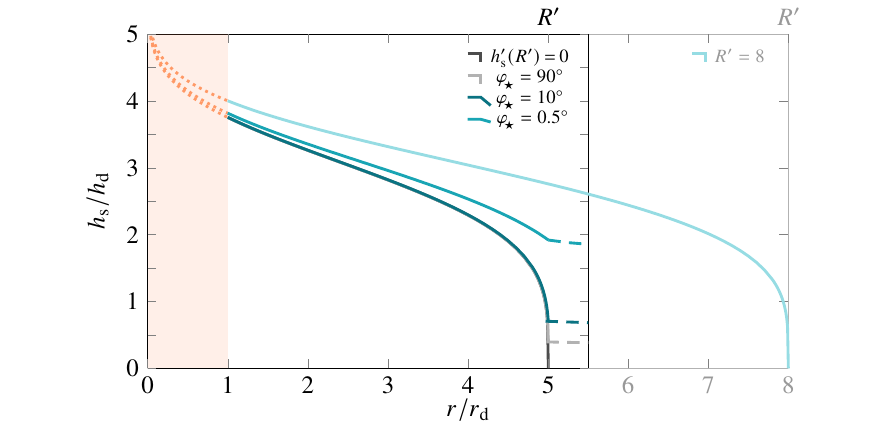}}
    \caption{Stationary film thickness profile computed using Eq.~\eqref{eq:cone-anal} over a flat stalagmite of radius $\Rsm' = 5$ in the main axes (\ie, the black ones). The curves are represented beyond a cut-off radius $r'=1$ to avoid accounting for the divergence of the solution in $r'\rightarrow0$. The curves in the main axes correspond to various values of the outer cone angle $\phiStar$,  used as a boundary condition: $\phiStar = \SI{90}{\degree}$ (light grey), $\SI{10}{\degree}$ (dark blue), and $\SI{0.5}{\degree}$ (blue).  The dashed lines of the same colours, in the continuity of the solid lines, show the corresponding approximated film thickness $\hfilmStnry' \sim (r')^{-1/3}$ for $r' \ge \Rsm'$ in each case (see Eq.~\eqref{eq:hsparabdimless}). The dark grey line represents the limiting case $\hfilmStnry'(\Rsm') = 0$ for $R' = \SI{5}{}$. The secondary part of the graph (\ie, with light grey axes), for $r \ge \SI{5.5}{\centi\meter}$, shows another profile with the same parameters and $\hfilmStnry(\Rsm) = 0$, for $\Rsm = \SI{8}{\centi\meter}$. The orange dotted lines and shaded area represent the virtual extension of the film thickness profiles for $r < 1$ for all the curves represented in the main and secondary axes. Primes denote nondimensional variables where explicit normalisation is absent.}
    \label{fig:hFilm-stationary-anal}
    \end{figure}

For flat stalagmites, $\shapeFac=0$ and $\smAng = 0$ $\forall \xi$. The derivatives $\dxi\cdot$ thus reduce to $\dr\cdot$ and Eq.~(\ref{eq:TimeIndep}) becomes
\begin{equation} \label{eq:hsdiffflat}
    - \dfrac{g}{3\nu} \hfilmStnry^3 \left( \dr \hfilmStnry \right) = \dfrac{\Vdrop}{2\pi r \tdrip}\, .
\end{equation}
This translates in nondimensional form to
\begin{equation} \label{eq:hsdiffflat}
- r' h'^3_{\rm s} \left( \partial_{r'} \hfilmStnry' \right) = \frac{3 \Vdrop'}{4 \tdrip'} \, ,
\end{equation}
where we used the characteristic scales introduced in Sec.~\ref{sec:charac-scales}, and included the same nondimensional drop volume as in the numerical resolution, \ie, $\Vdrop = \Vdrop' (\pi \rdrop^2 \hdrop/2)$ (see Sec.~\ref{sec:chap3-dropAdd}). 
Integrating the above equation yields
\begin{equation}
\hfilmStnry' = \left[  \frac{3 \Vdrop'}{\tdrip'} \ln \left( \frac{\Rsm'}{r'} \right)+ h'^4_{\rm s}(\Rsm') \right]^{1/4} ,
\end{equation}
where $\hfilmStnry'(\Rsm')$ is the value of the film thickness at the stalagmite edge, in $r'=\Rsm'$. Using $\hfilmStnry'(\Rsm') = 0$ would correspond to the film draining instantly in an artificial sink, the stalagmite outer wall (for $r' > \Rsm'$) remaining dry. While mathematically convenient, this scenario is not physically realistic. Stalagmites with a flat top are always surrounded by inclined walls, yielding residual drainage and a non-zero film thickness at the edge of the flat region in $r' = \Rsm'$. We approximate the outer region $r' > \Rsm'$ by a cone of constant opening angle $\phiStar$ over which the drained liquid film reaches a thickness $\hfilmStnry'(\Rsm')$ given by substituting $r' = \Rsm'$ and $\smAng = \phiStar$ in Eq.~(\ref{eq:hsparabdimless}) which is introduced in the following section. The film thickness on the flat top is therefore given by
\begin{equation} 
\hfilmStnry' = \left[  \frac{3 \Vdrop'}{\tdrip'} \ln \left( \frac{\Rsm'}{r'} \right)+  \left(  \frac{3 h_d}{4 r_d} \frac{V'_d}{\tdrip' \Rsm' \sin \phiStar} \right)^{4/3} \right]^{1/4}. \label{eq:cone-anal}
\end{equation}
The profile $\hfilmStnry'(r')$ is plotted in Fig.~\ref{fig:hFilm-stationary-anal} for various values of $\phiStar$. 
The opening angle of the cone at $r' > \Rsm'$ has a negligible influence on  $\hfilmStnry'$ at $r' < \Rsm'$ as long as $\phiStar \geq 10^{\circ}$. This verifies that the value $\phiStar = 45^{\circ}$ used in the simulations is sufficiently large to ensure that the outer boundary does not influence the stationary film profile. Equation~\eqref{eq:cone-anal} predicts an unphysical divergence of $\hfilmStnry'$ in $r' \rightarrow 0$, which we represented by the orange area and dashed line portions in Fig.~\ref{fig:hFilm-stationary-anal}, up to a cut-off radius $r' = 1$. Nevertheless, for sufficiently small $r'$, the first term in the bracket of Eq.~\eqref{eq:cone-anal} dominates the second, and $\hfilmStnry'$ becomes proportional to $(\tdrip')^{-1/4}$. 

The above scaling law is in agreement with the numerical simulations reported in Fig.~\ref{fig:hS_num}\;(a) for $\hfilmStnry'(r' \rightarrow 0)$. However, the simulations also reveal a proportionality of $\hfilmStnry'(r' \rightarrow 0)$ to $(\Rsm')^{1/4}$ that Eq.~(\ref{eq:cone-anal}) does not capture. This dependence can be retrieved by scaling $\partial_{r'} \hfilmStnry' \sim \hfilmStnry' / R$ in Eq.~(\ref{eq:hsdiffflat}), which then yields
\begin{equation} 
    \hfilmStnry' \simeq \left( \frac{3 \Rsm' \Vdrop'}{4 r' \tdrip'} \right)^{1/4}. \label{eq:hsflatdimless}
\end{equation} 
Although this scaling law is faithful to the dependence of $\hfilmStnry'(r' \rightarrow 0)$ in $\Rsm'$ and $\tdrip'$ observed in Fig.~\ref{fig:hS_num}\;(a), it still fails at producing a non-divergent solution in $r' \rightarrow 0$. Nevertheless, the meaning of the solution in $r' < 1$ should be questioned anyway since the liquid film is fed by an inflow that is physically of width $r' = 1$.

\subsubsection{Parabolic stalagmite}

For sufficiently curved stalagmites with $\shapeFac > 0$, there exist radial positions for which $\tan \smAng \gg | \dxi \hfilmStnry|$. In such conditions, Eq.~(\ref{eq:TimeIndep}) reduces to
\begin{equation} \label{eq:hsparab}
h^3_{\rm s} = \frac{3 \nu \Vdrop}{2 \pi g \tdrip r \sin \smAng}.
\end{equation}
The condition $\tan \smAng \gg | \dxi \hfilmStnry |$ can be made explicit by first differentiating this equation with respect to $\xi$: 
\begin{equation}
    3 \hfilmStnry^2 \left( \dxi \hfilmStnry\right) = \frac{3 \nu \Vdrop \cos{\smAng}}{2 \pi g \tdrip} \partial_r \left( \frac{1}{r \sin \smAng} \right).
\end{equation}
For a parabolic stalagmite of elevation $\smelv = - \shapeFac r^2$, $\tan \smAng = 2 \shapeFac r$ and 
\begin{equation}
\partial_r \left( \frac{1}{r \sin \smAng} \right) = - \frac{1 + \cos^2 \smAng}{r^2 \sin \smAng} \, .
\end{equation} 
We thus obtain
\begin{eqnarray}
    \frac{|\dxi \hfilmStnry|}{\tan \smAng} &=& \left( \frac{8 \nu \Vdrop \shapeFac^4}{9 \pi g \tdrip} \right)^{1/3} \frac{\cos^{10/3}{\smAng} \left(1 + \cos^{2}{\smAng} \right)}{\sin^{8/3} \smAng} \nonumber \\
&=&  \left( \frac{4 \Vdrop' \shapeFac'^4}{9 \tdrip'} \right)^{1/3} \left(\frac{h_{\rm d}}{r_{\rm d}} \right)^{8/3} \frac{\cos^{10/3}{\smAng} \left(1 + \cos^{2}{\smAng} \right)}{\sin^{8/3}{\smAng}} \, ,
\end{eqnarray}
where all the terms were nondimensionalised. 
For $\Vdrop' = 1$, $\tdrip' = 1$ and $\shapeFac' = 1$, the condition $\tan \smAng \gg  | \dxi \hfilmStnry|$ is valid for $\smAng \gg \SI{0.011}{\radian}$ (or $\SI{0.63}{\degree}$). For similar parameter values, $\smAng \ll 1$, $\sin \smAng \simeq \smAng$ and $\cos \smAng \simeq 1$, so the validity region of Eq.~(\ref{eq:hsparab}) is
\begin{equation}
    \smAng \gg \left( \frac{32 \Vdrop'}{9 \tdrip'} \right)^{1/8} \left( \frac{h_{\rm d}}{r_{\rm d}} \right) \left( \shapeFac' \right)^{1/2}.
\end{equation} 
Moreover, Eq.~(\ref{eq:hsparab}) translates in nondimensional form as
\begin{equation} \label{eq:hsparabdimless}
\hfilmStnry' = \left( \frac{3 h_{\rm d}}{4 r_{\rm d}} \frac{\Vdrop'}{\tdrip' r' \sin \smAng} \right)^{1/3} \, . 
\end{equation}
This scaling law rationalises the proportionality between $\hfilmStnry'$ and $(\tdrip')^{-1/3}$ observed far from the stalagmite centreline in the simulations reported in Fig.~\ref{fig:nooneb}\;(b). With the further assumption that $\sin \smAng \simeq \tan \smAng$ at small to moderate $\smAng$, we finally obtain
\begin{equation} \label{eq:hsparabdimless2}
\hfilmStnry' \simeq \left( \frac{3 \Vdrop'}{8 \tdrip' \shapeFac' r'^2} \right)^{1/3} \, ,
\end{equation}
The thickness-dominated and drainage-dominated mechanisms over flat and parabolic geometries, respectively, thus yield distinct dependences of the stationary film thickness on drop inflow. We no examine how this stationary film responds to the perturbations caused by the discrete drop inflow.

\subsection{Time evolution of a film thickness perturbation}

As observed in Figs.~\ref{fig:hs_r}, \ref{fig:heatMap} and \ref{fig:noone} from Sec.~\ref{subsec:num-examples}, the perturbation of the film thickness caused by the addition of a drop is simultaneously damped, diffused and advected outward. This combination of phenomena can be approached by considering the evolution of a small perturbation $\varepsilon$ of the time-independent film thickness, \ie,
$
h(\xi, t) \simeq \hfilmStnry(\xi) + \varepsilon (\xi, t)$, with $|\varepsilon| \ll \hfilmStnry. 
$
The drainage equation~(\ref{eq:drainage}) is then linearised in $\varepsilon$, which leads to
\begin{equation}
    \dt \varepsilon = \frac{1}{r} \dxi \left[ \frac{g h^2_{\rm s} r \cos{\smAng}}{\nu} \left( \dxi \hfilmStnry - \tan \smAng \right) \varepsilon + \frac{g h^3_{\rm s} r \cos \smAng}{3 \nu} \dxi \varepsilon \right] \, . 
\end{equation}
Assuming a localised perturbation for which $| r \partial_\xi \varepsilon | \gg \varepsilon$, the linearised equation becomes
\begin{equation}
    \dt \varepsilon + U \dxi \varepsilon = D \dxxi \varepsilon \, ,
\end{equation}
where $U$ and $D$ are functions of $\xi$, given by 
\begin{equation}
U = \frac{g h^2_{\rm s} \cos \smAng}{\nu} \left( \tan \smAng - 2 \, \dxi \hfilmStnry  \right) \, , \qquad D = \frac{g h^3_{\rm s} \cos \smAng}{3 \nu} \, . 
\end{equation}
The perturbation therefore obeys an advection-diffusion equation, with $U$ and $D$ being interpreted as the effective propagation velocity and diffusion coefficient. Close to the stalagmite centreline, $\smAng \ll 1$ so we can approximate $\sin \smAng \simeq \tan \smAng = 2 \shapeFac r$. According to Eq.~(\ref{eq:hsparab}), $\hfilmStnry$ is proportional to $\shapeFac^{-1/3}$ in this case. This makes $U$ proportional to $\shapeFac^{1/3}$, and $D$ proportional to $\shapeFac^{-1}$. The local  Péclet number that can be defined from there, $\Pe = U L / D$, compares advection to diffusion at a characteristic distance $L$. This Péclet number is then proportional to $\shapeFac^{4/3}$ for a fixed $L$. At sufficiently large $\shapeFac$ for which $\Pe \gg 1$, the perturbation should mostly be advected rather than diffused. This can be witnessed in Figs.~\ref{fig:hs_r} and \ref{fig:heatMap}, where perturbations are mostly diffused for $\shapeFac' \lesssim 1$, while they are mostly advected in the strongly curved case of $\shapeFac' = 10$. 

In the advection-dominated case, $\tan \smAng \gg |\dxi \hfilmStnry|$ and $\cos \smAng \simeq 1$. Substituting the scaling law of $\hfilmStnry$ from Eq.~\eqref{eq:hsparab} in the above expression of the perturbation velocity $U$ gives
\begin{equation}
U = \left( \frac{9 g V_{\rm d}^2 \sin \smAng}{4 \pi^2 \nu \tdrip^2 r^2 } \right)^{1/3} \, ,
\end{equation}
or, in nondimensional terms,
\begin{equation}
U' = \left( \frac{t_{\rm d}}{r_{\rm d}} \right) U = \left( \frac{9 \shapeFac' V'^2_{\rm d}}{8 t'^2_0 r'} \right)^{1/3}. 
\end{equation}
The perturbation at position $r'_{\rm p}$ will then travel according to the following differential equation:
\begin{equation}
\frac{{\rm d}r'_{\rm p}}{{\rm d} t'} = \left. \vphantom{\dfrac{a}{a}} U'\right|_{r' = r'_{\rm p}} \, .
\end{equation}
Integrating this with an initial condition $r'_{\rm p} = 1$ in $t' = 0$, we find 
\begin{equation}
r'_{\rm p} = \left[ \left( \frac{8 \shapeFac' V'^2_{\rm d}}{3 t'^2_0} \right)^{1/3} t' + 1\right]^{3/4}.
\end{equation}
This prediction is in good agreement with the observed propagation of drop-induced perturbations in Fig.~\ref{fig:heatMap}\;(d), for which we had numerically found that the front evolves as $(r')^{4/3}$ within the domain of the heat map. Having established the behaviour of a local perturbation, we now turn to the global unsteady drainage. 

\subsection{Unsteady drainage on a parabolic stalagmite in the absence of inflow}

When the drip inflow is low (namely, in the limit of large dripping period), the film may significantly drain between successive impacts. As formalised above, the perturbations associated with the drop impacts in the film are primarily advected. To predict the film thickness evolution in this case, Eqs.~\eqref{eq:model-conservativeCurvilinear-q} and \eqref{eq:model-hOfT} are combined into a single differential equation for $h(\xi,t)$: 
\begin{equation} \label{eq:drainage}
    \dt h = \frac{1}{r} \dxi \left[ \frac{g r h^3}{3 \nu} \left( \left( \dxi h \right) \cos{\smAng} - \sin{\smAng} \right) \right]. 
\end{equation}
Under the condition $\tan \smAng \gg |\dxi h|$, \ie, for curved profiles, the equation above becomes
\begin{equation} \label{eq:huparabdiff}
    \dt h = - \frac{1}{r} \dxi \left( \frac{g r h^3 \sin \smAng}{3 \nu} \right).
\end{equation}
Close to the stalagmite centreline, $\smAng \ll 1$ so we can approximate $\sin \smAng \simeq \tan \smAng = 2 \shapeFac r$, and $\dxi \cdot \simeq \dr \cdot$. We approximate the time-dependent solution using the following Taylor expansion: $h = h_{\rm u} + \alpha r^n$, with $n>1$ since $\dr h = 0$ in $r = 0$ by symmetry. The notation $h_{\rm u}$ refers to the unsteady film thickness. Substituting this Ansatz in Eq.~\eqref{eq:huparabdiff} and considering the limit where $r \rightarrow 0$ yields
\begin{equation}
    \dt h_{\rm u} \simeq - \frac{4}{3} \frac{g h_{\rm u}^3 \shapeFac}{\nu} \, .
\end{equation}
Integrating this relation over time and considering the film thickness to be $h_{{\rm u},0}$ at $t = 0$, we further find that
\begin{equation}
    h^{-2}_{\rm u} \simeq \frac{8 g \shapeFac}{3 \nu} t + h^{-2}_{{\rm u},0} \, .
\end{equation}
At times verifying
\begin{equation} \label{eq:huparabcrit}
t = \tdrip \gg \frac{3 \nu}{8 g \shapeFac h^2_{{\rm u}, 0}} \, ,
\end{equation}
the stationary film thickness $h_{\rm u}$ in $r = 0$ is approximated by
\begin{equation}
h_{\rm u} \simeq \left(  \frac{8 g \shapeFac \tdrip}{3 \nu}  \right)^{-1/2} \, ,
\end{equation}
or, in nondimensional form, 
\begin{equation} \label{eq:huparab}
h'_{\rm u} \simeq \left(  \frac{8 \shapeFac' \tdrip'}{3}  \right)^{-1/2} \, .
\end{equation}
This scaling law is observed in the numerical simulations of Figs.~\ref{fig:hS_num}\;(b) and \ref{fig:nooneb}\;(b). Assuming that the initial film thickness is larger than $h_{\rm d}$, the criterion~(\ref{eq:huparabcrit}) becomes, in nondimensional form, 
\begin{equation}
\tdrip' \shapeFac' \gg \frac{3}{8} \, ,
\end{equation}
which is equivalent to saying that $h' \ll 1$. This criterion is also verified in Fig.~\ref{fig:hS_num}\;(b), where the regime in $h' \sim (\shapeFac')^{-1/2}$ is only valid at sufficiently large curvatures $\shapeFac' > 1$. 

The unsteady film thickness $h'_{\rm u}$ given by Eq.~(\ref{eq:huparab}) becomes smaller than the steady film thickness $\hfilmStnry'$ given by Eq.~\eqref{eq:hsparabdimless2} for a parabolic case when 
\begin{equation} \label{eq:rus}
r' < \rStnry'  = \left( \frac{8 \shapeFac' \tdrip' V'^2_{\rm d}}{3} \right)^{1/4}.
\end{equation}
The above relation suggests that unsteady drainage is more efficient close to the centreline, where the steady drainage solution produced a divergent solution. 
One may consider $\rStnry'$ as the aforementioned limit position surrounding a central, unsteady region where the discreteness of drop impacts influences the drainage (see Sec.~\ref{subsec:num-examples}). Beyond this region, the unsteadiness of the inflow is smoothed out to produce a time-independent thickness profile. 
For instance, using the values $\tdrip' = 1$, $\Vdrop' = 1$ and $\shapeFac' = 0.5$ of Fig.~\ref{fig:flat-vs-incl-example}\;(f), we obtain $\rStnry' \simeq 1.1$, which matches well the radial distance beyond which the time-variations of $h$ between two successive impacts become negligible. Similarly, in the configurations of Fig.~\ref{fig:heatMap}\;(b-d) with $\tdrip' = 1$, $\Vdrop' = 1$ and  $\shapeFac' = 0.1$ (resp. $\shapeFac' = 1$, $\shapeFac' = 10$),  Eq.~(\ref{eq:rus}) predicts $\rStnry' = 0.7$ (resp. $\rStnry' =1.3$, $\rStnry' =2.3$). The heat maps shown in Fig.~\ref{fig:heatMap} confirm that time variations are indeed mostly confined in $r' \lesssim \rStnry'$. 
We further infer from Eq.~\eqref{eq:rus} that there exists a corresponding dripping period $t'_{0,{\rm u/s}}$ delimiting the unsteady/steady regions at fixed radial position, \eg, for $r' \simeq 1$. This yields
\begin{equation} \label{eq:dripRegion}
t'_{0,{\rm u/s}} = \frac{3}{8 \shapeFac' V'^2_{\rm d}} \, .
\end{equation}
Hence, the film thickness will be in the steady regime (resp. unsteady regime) for dripping periods $\tdrip' \lesssim t'_{0,{\rm u/s}}$ (resp. $\tdrip' \gtrsim t'_{0,{\rm u/s}}$).

\subsection{Transient propagation front}

We finally come back to the early transient dynamics of the film propagation over the stalagmite, from when the first drops are added. The system of Eqs.~\eqref{eq:model-conservativeCurvilinear-q} and \eqref{eq:model-hOfT} is hyperbolic at the liquid front position even though it becomes parabolic within the film (see Sec.~\ref{subsec:nature-gov-eqs}). Hence, it admits a similarity solution within this film, valid until the position of the moving front $\lFront$, as derived by~\cite{Huppert1982} in radial coordinates. Two particular cases were considered and correspond to limiting behaviours in our system covered in Sec.~\ref{subsec:num-examples}: (i) a known liquid volume spreading under the form of a puddle over an initially dried horizontal surface (\ie, a single drop), and (ii) a known flow rate continuously feeding this puddle (\ie, a continuous inflow, or jet). If the volume of the puddle is equal to,~\eg, the volume of a drop~$\vold$ in the first case, and if the flow rate feeding the puddle is~$\Qdrip$ in the second one, the puddle advances over the surface by covering a disk of radius $\rFront(t)$, given by
\begin{equation}
    \left. \rFront \vphantom{\dfrac{a}{a}} \right|_{\rm flat} \simeq \left\{
        \begin{array}{lr}
        0.9 \left( \dfrac{g \vold^{3}}{3 \nu}\right)^{1/8} t^{1/8} \, \text{,} &\quad \text{(constant volume)}\, ,  \\
        0.7 \left( \dfrac{g \Qdrip^{3}}{3 \nu }\right)^{1/8} t^{1/2} \, \text{,} &\quad \text{(continuous inflow)}\, , 
    \end{array}
    \right. \quad \shapeFac = 0 \, . 
    \label{eq:huppert}
\end{equation}
We thus retrieve the same scalings as obtained numerically for a flat case. Replacing the continuous inflow $\Qdrip$ by its discrete equivalent, both equations actually yield the same scaling $\lFront \sim \tdrip^{1/4}$, reminiscent of the crossover between both asymptotic behaviours observed in Fig.~\ref{fig:front-propagation}\;(a). 
In a general shape framework, there is no known analytical solution to the unsteady case of Eqs.~\eqref{eq:model-conservativeCurvilinear-q} and \eqref{eq:model-hOfT}, although asymptotic solutions were formerly derived in this section. These latter could further be extended numerically for variables curvature~$\shapeFac$ in the constant volume and continuous inflow regimes, along with the case of intermediate discrete drop inflow in Fig.~\ref{fig:front-propagation}. 

The front position scalings obtained numerically differ from the scaling obtained for the extent of the unsteady region, $\rStnry$, that we reviewed in the previous section. The unsteady region position is time-independent even though it depends on the geometry and drop inflow considered. The front position $\lFront$, however, is time-dependent. It characterises the instantaneous extent of the film tip, whereas $\rStnry$ denotes the length separating the unsteady, drop-dominated region from the quasi-steady region where the film evolves as under a continuous inflow. Hence, when $\lFront > \rStnry$, the film behaves as if fed by a continuous inflow, and the corresponding scaling laws converge toward those obtained in the continuous limit. 
This suggests that at the scale of the front near vanishing film thickness, the propagation dynamics becomes insensitive to the origin of the flow forcing, \ie, to whether the drainage is thickness-dominated or inclination-dominated. This behaviour can be interpreted as a manifestation of the hyperbolic nature of the governing equations at the front, where advective transport dominates over local accumulation. 
This might also explain how the discrete drop additions do not influence the asymptotic behaviour of the front at large timescales, in contrast with, \eg, the stationary film thickness at the stalagmite centre. 
The front thus propagates as if fed by a continuous inflow regardless of the discrete nature of the drops once $\lFront > \rStnry$, in line with the observations made on Fig.~\ref{fig:front-propagation}.

\section{Experimental stationary film thickness}
    \label{sec:exp}

We now complete our theoretical and numerical analysis with experimental measurements of the stationary film thickness. We first review briefly the methods used to obtain these measurements, then compare them to the scalings obtained. All the measured quantities are dimensional, but will be presented in the same nondimensional framework as developed in Sec.~\ref{sec:charac-scales} for consistency with the numerical results.  

    \subsection{Methods}
    \label{subsec:exp-meas}

    We report in Sec.~\ref{subsec:exp-res} hereafter experimental measurements of the stationary film thickness as a function of the drop dripping period~$\tdrip$ for a total of 20 stalagmites, 12 of which were considered as having a parabolic profile from photographs, and the remaining ones presenting a completely flat cap. Details are provided in Tab.~\ref{tab:exp-meas} from App.~\ref{app:sm-shape}, which sums up the dimensional measurements of the drop dripping period $\tdrip$, the stationary film thickness $\hfilmStnry$, the radius $\Rsm$ or shape factor $\shapeFac$. Among these stalagmites, 19 were 
    in situ, and one was transported to a controlled lab setting, after it had been broken during construction work in Aven d'Orgnac (France). Its top is nevertheless undamaged. 
    The caves provided a great diversity of shapes and dripping periods. During the lab measurements, the drop inflow could be varied systematically for a fixed shape. The measurements come from various caves. 
    The range of radii for flat stalagmites comes from the scattering in the drop impact point position, which affects the stalagmite spatial extension and increases with the falling height of the drops. Their falling height varied on a large range (from $\lesssim \SI{1}{\meter}$ to $\gtrsim \SI{25}{\meter}$). The estimated shape factor $\shapeFac$ for most stalagmites remains quite small, between $\sim 0.05$ and $\sim 0.2$, \ie, most stalagmites do not present a steep parabolic profile. This does not stem from image truncation nor visual bias but rather from the common decoupling observed, at the stalagmite scale, of a body part of more or less constant width, surmounted by a cap exhibiting a very flat or parabolic-like profile.

    Measurements of the film thickness on cave stalagmites were obtained by a differential weighing technique, using a small piece of paper towel of known density and surface area (a $\sim\SI{2}{\centi\meter}$-side square, close to the crushing drop expansion diameter formed at impact, see Sec.~\ref{sec:charac-scales}) to collect water at the centre of the stalagmite. This leads to surface-averaged measurements, hence the size of the towel paper piece was carefully chosen as not too close to the stalagmite radius, but still large enough for collecting a sufficient weight of water and not induce large errors in the weighing measurements. Measurements were made after artificially stopping the drop inflow at several post-impact times. This allows to infer the initial film thickness of the curve obtained. Regarding the lab measurements, a Keyence confocal sensor CL-P030 was used for live recording of the film thickness as drops were successively added into the film. Drops were created using an appropriate piece of plastic tubing and a $\SI{2}{\milli\meter}$-radius Luer-Lock plastic adapter to recreate drops of radius close to caves ($\sim\SI{2.3}{\milli\meter}$) combined with a Watson Marlow peristaltic pump, in controlled high-humidity conditions. Splash was avoided at drop impact in the film by positioning the tip of the tube close to the film interface ($\sim \SI{4}{\centi\meter}$ above it). Measurements were taken as close as possible to the drop injection point ($\lesssim\SI{1}{\centi\meter}$) given the geometric constraint of the sensor. Examples of typical film thickness time series are presented in App.~\ref{app:sm-shape}.  Further details on the experimental protocols and time-resolved film measurements are provided in \cite{Parmentier2024}. 

    \subsection{Results and comparison}
    \label{subsec:exp-res}

    \begin{figure}
    \centerline{\includegraphics[width = \linewidth]{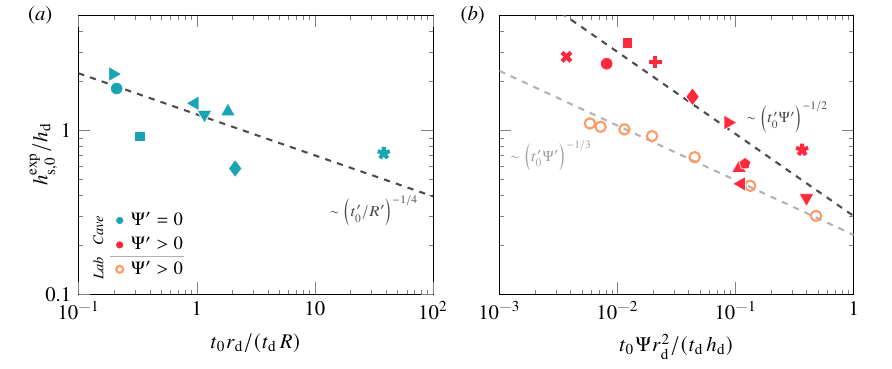}}
        \caption{
        Stationary film thickness from cave and lab measurements as a function of (a) $\tdrip'/\Rsm'$ for flat stalagmites, and (b) $\tdrip'\shapeFac'$ for parabolic stalagmites. Each stalagmite is associated with a different symbol described in Tab.~\ref{tab:exp-meas}. The scaling laws from Eqs.~\eqref{eq:hsflatdimless} ($\sim (\tdrip' / \Rsm')^{-1/4}$), \eqref{eq:hsparabdimless2} ($\sim (\tdrip'\shapeFac')^{-1/3}$), and Eq.~\eqref{eq:huparab} ($ \sim (\tdrip'\shapeFac')^{-1/2}$) are represented by dashed lines. Measured quantities from Tab.~\ref{tab:exp-meas} are nondimensionalised according to Sec.~\ref{sec:charac-scales}. Predictions come from the main text but correspond closely to the fits made. The legend from (a) applies to both panels. Primes denote nondimensional variables where explicit normalisation is absent.
        }
        \label{fig:hS_exp}
    \end{figure}

    The cave and lab measurements of stationary film thickness are reported in Fig.~\ref{fig:hS_exp}. They are either plotted as a function of $\tdrip'/\Rsm'$ in panel (a) showing flat stalagmites, or as a function of $\tdrip' \shapeFac'$ in panel (b)  for parabolic ones. Data were grouped according to the best estimated stalagmite shapes, with one symbol per stalagmite as detailed in Tab.~\ref{tab:exp-meas}. 
    Although also parabolic-like, the lab stalagmite was plotted separately to better visualise the effect of variable inflow on this particular stalagmite in panel (b). 
    The theoretical scalings derived in the nondimensional framework are overlaid as reference lines: $\sim(\tdrip' / \Rsm')^{-1/4}$ from Eq.~\eqref{eq:hsflatdimless} for the flat case in panel (a), and $\sim(\tdrip' \shapeFac')^{-1/3}$ from Eq.~\eqref{eq:hsparabdimless2}, and $\sim(\tdrip' \shapeFac')^{-1/2}$ from Eq.~\eqref{eq:huparab} for curved stalagmites in (b). These guide the comparison between experimental measurements and model predictions. 
    
    A good agreement is observed between the flat stalagmite data and the theoretical prediction in panel (a), with a fitted exponent of $-0.22$ from the measurements. For the parabolic-like stalagmites, the slope of $-1/2$ matches the cave data reasonably well in logarithmic scale, with an experimental fit yielding an exponent of $-0.46$ on the cave subset alone. In contrast, the laboratory stalagmite exhibits a milder dependence, with a fitted exponent of $-0.30$, consistent with drainage dominated by the local slope rather than by global unsteadiness. The measurements on the lab stalagmite were recorded within a short distance from the point of drop impact (typically $\lesssim \SI{1}{\centi\meter}$), where the local inclination of the surface likely governs the film flow. A slight bank angle of the sample may also have contributed to this effect. Lab data were also continuously recorded under a steady dripping inflow. In contrast, the cave data come from temporal measurements following the addition of a single drop, during which the inflow was temporarily interrupted. The stationary thickness was then reconstructed as the initial thickness of the measured curves. Consequently, the lab data should genuinely represent a film at stationary state, while the cave data are more representative of unsteady drainage without new drop additions and could rather be named $\hfilm_{\rm u}$. An approximate dripping period $t_{\rm 0, u/s}'$ delimiting unsteady/steady regimes at fixed shape is defined by Eq.~\eqref{eq:dripRegion} in Sec.~\ref{subsec:scalings}. Accounting for the larger drops in caves and the fact that most stalagmites have an associated shape factor $\shapeFac' \sim 0.05$, this dripping period reduces here to $\sim 0.3$. This corresponds roughly to the region where cave and lab measurements overlap in Fig.~\ref{fig:hS_exp}\;(b). Altogether, these results reinforce the existence of distinct drainage regimes in parabolic geometries: a local, slope-dominated regime at stationary state, and a global, unsteady regime where curvature and inflow intermittency jointly control the film evolution.

    \begin{figure}
    \centerline{\includegraphics[width = \linewidth]{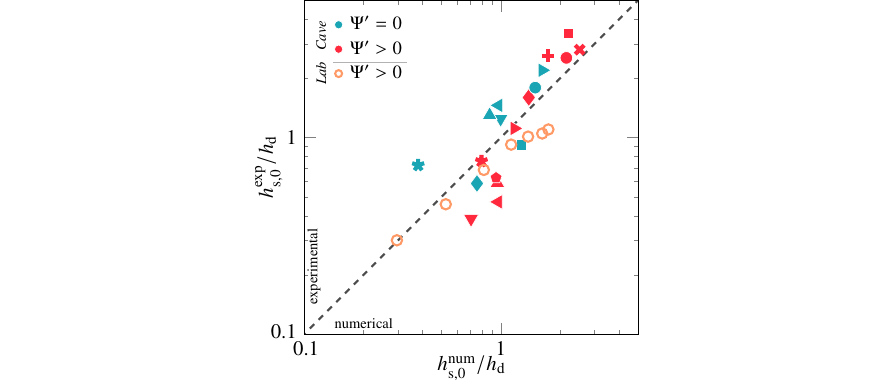}}
        \caption{
        Parity plot comparing numerical predictions (based on $\tdrip'$, $\Rsm'$, $\shapeFac'$) and experimental stationary film thickness $\hfilmStnryZ'$. The dashed line is the bisector line. Stalagmites were split into flat and parabolic groups, as indicated by the symbol colours. Each stalagmite was attributed a specific symbol, whose correspondence can be found in Tab.~\ref{tab:exp-meas} along with the measured values of $\Rsm'$ (radius for flat stalagmites) or $\shapeFac'$ (shape factor for parabolic ones). All quantities were nondimensionalised according to Sec.~\ref{sec:charac-scales}.}
        \label{fig:hS_parity}
    \end{figure}
    
    Finally, in Fig.~\ref{fig:hS_exp} (b), a parity plot compares the experimental measurements to their numerical counterpart. The input parameters of the simulation, namely $\shapeFac'$, $\Rsm'$, and $\tdrip'$, were calculated from the cave and lab measurements from Tab.~\ref{tab:exp-meas}. The numerical stationary film thickness was computed by rescaling the measured dripping periods using the drainage timescale~$\tdrain$ for consistency with the numerical results. The measured radii and shape factors were nondimensionalised using the former values of $\rdrop = \SI{1}{\centi\meter}$ and $\hdrop = \SI{100}{\micro\meter}$ according to Sec.~\ref{sec:charac-scales}. The overall trend of the parity plot exhibits a good agreement between experimental and numerical data, although some deviations are visible. The root mean square error is about $0.42$, with a relative error of about $\SI{31}{\percent}$ across all measurements, and a correlation coefficient of 0.69. These non-negligible discrepancies can be attributed to several limiting factors: (i) the accuracy of the cave measurements, based on a mechanical technique rather than an optical one, (ii) uncertainties in the shape factor estimation, and (iii) other effects (\eg, capillarity, impact point dispersal and splashing) omitted in the modelling. The shape factor in particular was estimated from stalagmite photographs possibly taken with a bank angle or an uneven lighting. Some stalagmites might also not have a perfectly axisymmetric shape or present irregularities that further complicate the shape factor estimation. Finally, the flat and parabolic approximations might not be best suited for all stalagmites.

    In summary, besides the above discussed limitations, the experimental measurements broadly support the theoretical and numerical predictions. They confirm the existence of the different drainage regimes identified, with a transition mostly influenced by geometry-driven drainage dynamics. We sum up the results hereafter.

\section{Discussion}
\label{sec:discussion}

The drainage of a thin film over a stalagmite of general shape results from the balance between gravity and viscosity, modulated by the local surface inclination. Within the lubrication framework, we analysed the film response to the combined effects of shape variability and discrete drop inflow. A summary of the main variables used throughout the study is provided in Tab.~\ref{tab:nomenclature}.

Once at stationary state, the film thickness over a fixed stalagmite shape is uniquely defined by the discrete drop inflow. For flat stalagmites, numerical and experimental results consistently follow the theoretical scaling $\hfilmStnry \sim \tdrip^{-1/4}$ for the stationary film thickness. For parabolic geometries, a local theoretical drainage balance rather predicts that the unsteady film thickness evolves as $\sim \tdrip^{-1/2}$ at the centreline. Away from the centreline, the stationary film thickness should follow $\hfilmStnry \sim \tdrip^{-1/3}$ because the local surface slope predominates, and the film evolves as if fed from a continuous inflow. The scalings from numerical simulations and experiments are consistent with the theoretical derivations. 
In particular, over parabolic stalagmites, the different protocols of measurements correspond to various expressions of the drainage at play. Pointwise lab measurements away from the dripping point over an inclined slope display a stationary behaviour, while drop surface-averaged cave measurements for an interrupted inflow capture the unsteady evolution of the film thickness. 

All these results underlie the coexistence of local and global effects, as well as steady/unsteady regions, which compare the effect of individually adding drops to the film to the unsteady drainage. 
The differences observed across shapes stem from the physical mechanisms at play. Over flat geometries, the film is driven by liquid accumulation and spreading over the entire surface whereas for parabolic stalagmites, the more efficient gravity-induced drainage prevents local build-up. 
The asymptotic dependence of the stationary film thickness on the stalagmite shape can also be theoretically derived. For flat cases, the stationary film thickness follows $\hfilmStnry \sim \Rsm^{1/4}$ at first order, whereas parabolic geometries unsurprisingly yield a stationary film thickness decreasing with increasingly curved profiles. 
The propagation of the drop-induced perturbations in the film is primarily driven by the different shapes as well. For flat stalagmites, the perturbations are damped within the film according to a diffusive process. Conversely, on curved surfaces, perturbations are more rapidly advected in a wave-like motion.

Additionally, the front propagation consistently follows a power-law scaling in both flat and parabolic geometries, although the relative exponents differ. There is nevertheless a shared scaling law $\lFront \sim t^{1/2}$ found for the front propagation resulting from various combinations of stalagmite shape and inflow regime considered, \ie, different driving mechanisms. Accordingly, once the time-dependent front position lies outside of the steady region, it behaves as in the continuous limit regardless of the discrete nature of the drops. This suggests that the front locally behaves as a self-similar propagating wave.

The present modelling relies on several simplifying assumptions. We focused solely on axisymmetric stalagmites with perfectly smooth surfaces, thereupon avoiding possible film thickness variations due to lateral, capillary or rugosity effects. Similarly, drops were assumed to impact the film at a central position without splash, a condition that we could recreate in a lab setting. Although limiting, the hypotheses regarding the absence of splash at impact and the non-scattered drop impact point position distribution remain representative of low-falling height impacts in caves. On average, drops falling from high cave ceilings still land around the centre of the stalagmite. The impact point dispersal can thus be viewed as an additive local noise broadening the stalagmite width and unsteady/steady region frontier at first order. The splash at impact lasting for a few milliseconds only induces a short-term dynamics that is not considered here. Additionally, the mixing between the drop and the film at the impact location should be considered when modelling the entire coupling of film drainage combined with ion precipitation to more accurately describe stalagmite growth. These simplifications clarified the fundamental drainage mechanisms at play, but there are likely secondary local effects not captured here. 

Overall, despite the simple measurement techniques used in caves, there is a reasonable correspondence between the theoretical scalings, and the numerical and experimental results. Each of them sheds light on different aspects of how the surface inclination and drop inflow affect gravity-driven drainage. 
Altogether, the stationary film thickness and unsteady drop propagation analysed in this study reveal the richness of a process as simple as the thin film drainage over stalagmites.

\begin{table}
    \begin{center}
    \def~{\hphantom{0}}
        \begin{tabular}{clc}
            Symbol & Meaning  & Dimensions \\[3pt]
            \hline
            $\nu$ & Kinematic viscosity of water & $\mathrm{L}^2\mathrm{T}^{-1}$\\
            $g$ & Gravitational acceleration & $\mathrm{L}\mathrm{T}^{-2}$ \\
            $r, z$ & Radial coordinates & $\mathrm{L}$ \\
            $\xi, \zeta$ & Curvilinear coordinate along/across the film thickness & $\mathrm{L}$ \\
            $\smelv$ & Stalagmite vertical elevation & $\mathrm{L}$ \\
            $\Rsm$ & Stalagmite radius/computational domain size & $\mathrm{L}$ \\
            $\shapeFac$ & Shape factor (curvature parameter) & $\mathrm{L}^{-1}$ \\
            $\smAng$ & Stalagmite local inclination & - \\
            $\phiStar$ & Cone opening angle at the boundary & - \\
            $p$ & Pressure in the film & $\mathrm{M}\mathrm{L}^{-1}\mathrm{T}^{-2}$ \\
            $u$ & Velocity in the film & $\mathrm{L}\mathrm{T}^{-1}$ \\
            $\flux$ & Integrated cross-sectional flux & $\mathrm{L}^2\mathrm{T}^{-1}$ \\
            $\Vdrop$ & Drop volume & $\mathrm{L}^3$ \\
            $\Qdrip$ & Drop volumetric inflow & $\mathrm{L}^3\mathrm{T}^{-1}$ \\
            $\tdrip$ & Drop dripping period & $\mathrm{T}$ \\
            $\hfilm$ & Film thickness & $\mathrm{L}$ \\
            $\hfilmStnry$ & Stationary film thickness & $\mathrm{L}$ \\
            $\hfilmStnryZ$ & Stationary film thickness at the centreline & $\mathrm{L}$ \\
            $\lFront$ & Film front position & $\mathrm{L}$ \\
            $\rStnry$ & Transition length between unsteady and steady regions & $\mathrm{L}$ \\
            $\tMax$ & Post-impact time at which maximum film thickness is reached & $\mathrm{T}$ \\
            \hline
            $\rdrop$ & Characteristic tangential length scale & $\mathrm{L}$ \\
            $\hdrop$ & Characteristic film thickness scale & $\mathrm{L}$ \\
            $\tdrain$ & Characteristic drainage transport timescale & $\mathrm{T}$ \\
        \end{tabular}
        \caption{Main symbols and dimensional quantities.}
        \label{tab:nomenclature}
    \end{center}
\end{table}

\section{Conclusions}

In this study, we have investigated the interplay between discrete drop inflow and substrate geometry on the drainage dynamics of thin films over stalagmites. Starting from Reynolds lubrication theory expressed in a curvilinear coordinate system to account for the variable stalagmite shapes, we derived temporal equations relating the local film thickness to the surface-parallel flux within the film. By completing these equations with the continuity and physically relevant boundary conditions, we could solve the non-linear problem obtained numerically. Combining a theoretical analysis and the numerical results, we identified the dominant drainage regimes governing the stationary film thickness evolution that we compared to experimental measurements taken in cave and in lab. 

The novelty in our modelling resides in the explicit inclusion of the discrete drop inflow, rather than a time-averaged inflow. This approach revealed that distinct regimes of steady or unsteady inflow appear at different locations and depending on the ratio between dripping period and a characteristic timescale related to the drainage outflow. These regimes are important for real stalagmites as the dripping period in caves naturally spans several orders of magnitude. The model also quantifies how the stalagmite shape directly affects the stationary film thickness and the attenuation of the perturbation caused by the discrete drops in the film. 
Ultimately, this framework offers a foundation for future predictions aiming at coupling flow and ion transport to mineral deposition for stalagmite growth modelling and palaeoclimate reconstruction.


\backsection[Acknowledgements]{This work was conducted at the University of Liège, Belgium. The authors gratefully thank the cave authorities from Clamouse cave, Aven d'Orgnac and La Salamandre for their warmful welcome (in particular, Amaury Guichon and Stéphane Tocino). They also ackowledge the help of Sophie Lejeune, Kevin Bulthuis and Fran\c{c}ois Bourges for participating in collecting the reported cave measurements. JP thanks Antonio Martinez Aguilera, Antoine Desiron, Mathieu Torfs and Olivier Verlaine for their precious help setting up the cave and lab experiments.}






\appendix

    \section{Experimental measurements}\label{app:sm-shape}

\begin{figure}
    \centerline{\includegraphics[width = \linewidth]{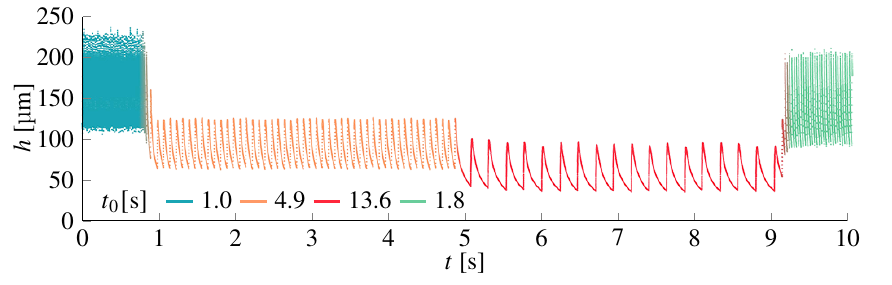}}
        \caption{Example of experimental film thickness taken over Lab01 as a function of time, for variable drop dripping period over time.}
        \label{fig:kd}
    \end{figure}

    In Tab.~\ref{tab:exp-meas}, we report experimental measurements of film thickness, in several caves from Southern France and in a more controlled lab setting. These measurements are presented in Fig.~\ref{fig:hS_exp}. In Fig.~\ref{fig:kd}, the evolution of the film thickness over time is shown over the stalagmite from the lab (Lab01). A succession of discrete drop impacts were made with a variable dripping period. The figure exemplifies how the film quickly adapts to changing conditions. It also shows how a given stalagmite shape and drop inflow uniquely define the film thickness with no memory effect. The lab stationary film measurements were averaged over a repetition of $\gtrsim 100$ impacts as those shown in Fig.~\ref{fig:kd}.

    \begin{table}
        \begin{center}
        \def~{\hphantom{0}}
        \centerline{\includegraphics[width = \linewidth]{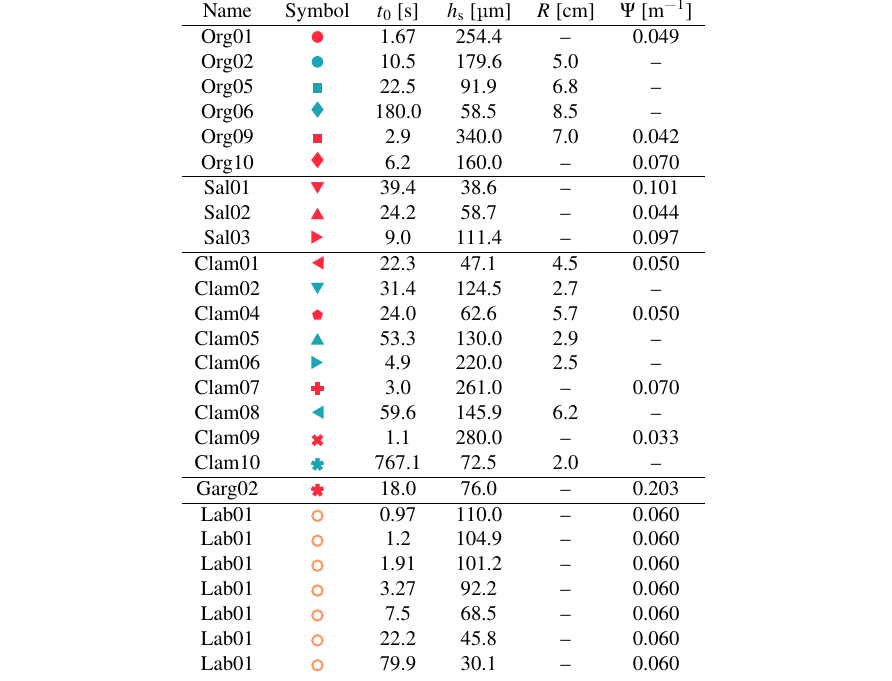}}
                \caption{Experimental measurements in cave and lab. Column 1 displays the name of the stalagmite from a southern France cave (numbering comes from previous studies): Org - \textit{Aven d'Orgnac}, Sal - \textit{La Salamandre}, Clam - \textit{Clamouse}, Garg - \textit{Gargas}, Lab - broken stalagmite from \textit{Aven d'Orgnac} used in the lab. Column 2 shows the associated symbol from Fig.~\ref{fig:hS_exp}. Columns 3 to 6 respectively represent the average drop dripping period (s.d. $\lesssim \SI{10}{\percent}$), the average stationary film thickness, the radius of flat stalagmites, and the shape factor of parabolic-like stalagmites. }
                \label{tab:exp-meas}
        \end{center}
    \end{table}

    \section{Numerical boundary condition at the centreline}
    \label{app:chap3-bc-num}

    We discuss here the implementation of the boundary condition derived at the centreline to close the system of equations in Secs.~\ref{subsec:nature-gov-eqs} and \ref{sec:front}. We express the next equations using nondimensional variables written without primes, in accordance with the numerical derivation from the main text (see Sec.~\ref{sec:theory}). First, we consider a discrete inflow of drops, then a laminar jet continuously replacing this drop inflow. At the outer boundary ($r = R$), a constant flux condition was imposed to replicate the effect of a steady outflow of liquid over a cone of constant opening angle. Tests performed with larger $R$ values confirmed that the resulting drainage behaviour at the centre is insensitive to this boundary. In principle, a Robin-type condition, relating $\flux(R)$ to the local film thickness, could provide a more general formulation, but this refinement lies beyond the scope of the present model.

        \subsection{Discrete drop inflow} \label{subsec:discrete}
    
        By mass conservation, the inflow of drops arriving in the first numerical cell of the domain is entirely transferred to the following cells. At the outer face of the first cell, $\hfilm_{\unDemi}^{n} = \left(\hfilm_0^{n} + \hfilm_1^{n}\right)/2$, and $\flux_{\unDemi}^{n} r_{\unDemi}$ becomes $\flux^{n}_{\unDemi} \Dxi/2$. The flux at the first cell outer face writes as
        \begin{equation}
            \flux_{\unDemi}^{n} = -\dfrac{1}{3} \left( \hfilm_{\unDemi}^{n} \right)^3 \left( \dfrac{\hfilm_1^{n} - \hfilm_0^{n}}{\Dxi} \right) \, .
        \end{equation}
        Using finite differences, the divergence term of Eq.~\eqref{eq:model-h-nd} becomes $2/\Dxi \left( 2 \, \flux_{\unDemi}^{n} r_{\unDemi}  / \left( \Dxi/2 \right) \right)$. The boundary condition at the first cell centre thus reads
        \begin{equation}
            \hfilm_0^{n+1} = \hfilm_0^{n} - \dfrac{ 4 \microSpace \Dt \, \flux_{\unDemi}^{n}}{\Dxi} \, .
        \end{equation}
    
        Regarding the outer limit of the stalagmite, we use the approximate condition derived in Sec.~\ref{subsec:scalings} relative to a cone with constant opening~$\phiStar$. The opening angle can be considered as equal to~$\smAng_{I - \unDemi}$ for parabolic or inclined profiles, while a value for~$\phiStar = \SI{45}{\degree}$ is imposed in flat cases. 
        The flux at the inner face of the last cell is given by matching the corresponding flux that would be obtained on the outer cone, leading to
        \begin{equation}
            \flux_{I - \unDemi}^{n} = \dfrac{\left( \hfilm^{n}_{I - \unDemi} \right)^{3}}{3} \left(\dfrac{\rdrop}{\hdrop}\right) \sin{\smAng_{I - \unDemi}} \, \text{,}
        \end{equation}
        with the film thickness at the inner face approximated by $\hfilm_{I - \unDemi}^{n} = \left( \hfilm_{I}^{n} - \hfilm_{I - 1}^{n} \right)/2$.  
        The corresponding update for the film thickness at the last cell centre reads 
        \begin{equation}
            \hfilm^{n+1}_{I} = \hfilm^{n}_{I} - \dfrac{2 \microSpace \Dt \left( 2 R -  \Dxi \microSpace \cos{\smAng_{I - \unDemi}} \right) \miniSpace \flux_{I - \unDemi}^{n} }{R \miniSpace \Dxi} \, .
        \end{equation}

        \subsection{Constant pointwise inflow}
        
        We now consider the case of a constant inflow for the front propagation calculations. This inflow can be viewed as either an infinitesimally thin jet of undefined height arriving at the centre of the stalagmite, or, equivalently, as a column jet of arbitrary width. By fixing the inflow $\qin$ and width $\rnought$ of the column of liquid, we can approximate the height $\hnought$ of the film at the radial position corresponding to the liquid column outer radius.  
        It is related to the corresponding continuous inflow through $\qin = -\rnought\hnought/(4\tdrip)$. Numerically, this inflo becomes
        \begin{equation}
            \qin = \dfrac{-1}{4 \Dt \hspace{0.75pt} \ndrip} \, ,
        \end{equation}
        with $\ndrip$ the number of added drops. 
        At the column outer radius, the stalagmite underneath surface may be considered as relatively flat, such that $\smAng \simeq 0$ locally. If the liquid column height is equal to~$\hnought$ and the film thickness at a distance~$\Dxi$ from this liquid column to~$\hun$, the thickness gradient within Eq.~\eqref{eq:model-q-nd} can be approximated as $\left( \hnought - \hun \right)/\Dxi$. Hence, we can impose a numerical, nondimensionalised boundary condition at the outer radius of the liquid column to find the thickness relative to the inflow as follows:
        \begin{equation}
            \hun = \hnought - \dfrac{3 \Dxi \qin}{\hnought^3} \, .
        \end{equation}
        The value of the column height at the radius~$\rnought$ comes from the approximation made for the drop volume brought into the film over a time~$\tdrip$: $\hnought = 1/\rnought^2$. The value of $\hnought$ could also be set by considering,~\eg, a cylinder spread over a radius~$\Dxi$. However, choosing that $\hnought$ is spread over $\rnought = 1$ leads to better numerical stability.

\bibliographystyle{jfm}
\bibliography{jfm}

\end{document}